\documentclass[a4paper,11pt]{article}
\usepackage{jheppub}

\usepackage{mathtools}
\usepackage{amssymb}
\usepackage{graphicx}
\usepackage{xcolor}
\usepackage{booktabs}
\usepackage{array}
\usepackage{comment}
\usepackage{microtype}

\graphicspath{{./}{./figures/}}

\newcommand{\dd}{\mathrm{d}}
\newcommand{\vev}[1]{\left\langle #1\right\rangle}
\newcommand{\vprop}{\bar v_{\mathrm{prop}}}

\newcommand{\rhodyn}{\rho_s^{\mathrm{dyn}}}
\newcommand{\AFB}{A_{\mathrm{FB}}}
\newcommand{\As}{\mathcal{A}_s}
\newcommand{\rhobg}{\rho_{\mathrm{bg}}^{\mathrm{dyn}}}

\newcommand{\safeincludegraphics}[2][]{%
  \IfFileExists{#2}{\includegraphics[#1]{#2}}{%
    \fbox{\begin{minipage}[c][3.4cm][c]{0.90\linewidth}
      \centering
      \textbf{Missing figure}\\[0.4em]
      \texttt{\detokenize{#2}}
    \end{minipage}}%
  }%
}

\title{\boldmath Fluctuation-Induced Backreaction on Bubble-Wall Dynamics in Cosmological First-Order Phase Transitions}

\author[a,b,c]{Dongdong Wei}
\author[c,b,a]{and Zong-Kuan Guo}

\affiliation[a]{School of Fundamental Physics and Mathematical Sciences,
Hangzhou Institute for Advanced Study, University of Chinese Academy of
Sciences, Hangzhou 310024, China}
\affiliation[b]{School of Physical Sciences, University of Chinese Academy of
Sciences, No. 19A Yuquan Road, Beijing 100049, China}
\affiliation[c]{Institute of Theoretical Physics, Chinese Academy of Sciences,
Beijing 100190, China}

\emailAdd{weidd@ucas.ac.cn}
\emailAdd{guozk@itp.ac.cn}

\abstract{
We study bubble-wall dynamics in cosmological first-order phase transitions in a
two-scalar-field model, where the wall is formed by the scalar field $\phi$ and
an additional real scalar field $s$ is coupled to it through a portal
interaction. We evolve the coupled classical field equations on the lattice and
consider a prescribed classical stochastic background of $s$ fluctuations at
the onset of the phase transition. Through the portal interaction, this patchy
background intermittently modulates the local driving pressure on the wall. The
wall consequently undergoes alternating episodes of acceleration and
deceleration, with a smaller finite-time averaged propagation rate than in the
decoupled limit. A wall-frame analysis of the dynamical $s$-sector energy reveals a
coupling-dependent crossover from excess energy deposited behind the wall at
weak coupling to a front-side bias at strong coupling. These results demonstrate
an additional scalar-field source of bubble-wall backreaction and may affect
gravitational-wave and baryogenesis predictions.}


\begin{document}
\maketitle
\flushbottom

\section{Introduction}
\label{sec:introduction}

Cosmological first-order phase transitions proceed through the nucleation and
expansion of bubbles of the broken phase. They are well-motivated sources of a
stochastic gravitational-wave background and are central ingredients in many
baryogenesis scenarios. Their real-time dynamics are therefore relevant to both
particle cosmology and future space-based gravitational-wave observatories such
as LISA, Taiji, and TianQin
\cite{LISA:2017pwj,LISA:2024hlh,Hu:2017mde,Luo:2021qji,
TianQin:2020hid,TianQin:2015yph}.

A key quantity is the bubble-wall velocity. It affects the efficiency and
characteristic lifetime of acoustic gravitational-wave sources and the
partition of released vacuum energy between scalar-field energy and bulk plasma
motion \cite{Hindmarsh:2013xza,Espinosa:2010hh,Caprini:2019egz,
Wang:2020jrd}. For electroweak-scale transitions, it also controls the transport
of CP-violating particle-number asymmetries near and in front of the wall
\cite{Morrissey:2012db,Cline:2020jre}. Bubble expansion is driven by the
free-energy difference between the phases and opposed by the backreaction of
the surrounding medium and other coupled degrees of freedom. If the
backreaction cannot balance the driving pressure, the wall may continue to gain
kinetic energy and approach an ultrarelativistic regime
\cite{Bodeker:2009qy}. Such rapid motion can suppress charge diffusion relevant
to electroweak baryogenesis \cite{Cline:2020jre} and alter the energy budget
entering gravitational-wave predictions
\cite{Caprini:2019egz,Espinosa:2010hh,Hindmarsh:2013xza,Leitao:2015fmj}.

Several complementary methods have been developed to determine or constrain
bubble-wall propagation. Semiclassical transport calculations obtain an
effective friction from departures of particle distributions from equilibrium
in a moving wall background \cite{Moore:1995ua}. Hydrodynamic backreaction can
also obstruct acceleration through compression heating and the redistribution
of released energy \cite{Konstandin:2010cd,Krajewski:2024gma}. In scalar--fluid
simulations, the phase-transition field is evolved together with a relativistic
fluid, and their interaction is commonly represented by a phenomenological
damping term \cite{Kurki-Suonio:1995yaf,Ignatius:1993qn,
Hindmarsh:2013xza}.

In this work we investigate a complementary mechanism that does not introduce a
phenomenological fluid-friction coefficient. We couple the scalar field $\phi$
that drives the transition to an additional scalar field $s$ carrying a
spatially inhomogeneous classical background. Through the portal interaction,
the local value of $s$ changes the free-energy difference sampled by the wall.
The driving pressure therefore becomes patch dependent, producing alternating
episodes of acceleration, deceleration, and, at sufficiently strong coupling,
transient blocking.

To isolate the backreaction induced by the additional scalar field, we
initialize $s$ with a classical stochastic background while neglecting the
subdominant stochastic component of $\phi$. This asymmetric initialization is a
controlled choice of initial conditions rather than a general assumption about
cosmological phase transitions.

The roles of the two simulation geometries are separated throughout the paper.
The $3+1$ dimensional simulations establish that the mechanism persists for an
expanding spherical bubble in a genuinely inhomogeneous background, while the
$1+1$ dimensional simulations provide the high-statistics coupling scan,
wall-normal energy diagnostics, and numerical convergence tests. The remainder
of the paper is organized as follows. Section~\ref{sec:setup} presents the
model, initial conditions, lattice evolution, and observables.
Section~\ref{sec:dynamics} combines the three- and one-dimensional results on
wall propagation. Section~\ref{sec:energy} studies the redistribution of energy
in the additional field, and appendix~\ref{app:numerical_validation} collects the main
numerical checks. We summarize the physical interpretation, limitations, and
conclusions in section~\ref{sec:summary}. Technical details are given in
appendix~\ref{app:numerical}.

\section{Model}
\label{sec:setup}

\subsection{Two-field model}
\label{subsec:model_initial}

We consider two real scalar fields. The field $\phi$ drives the first-order
phase transition, and its interpolation between the false- and true-vacuum
values forms the bubble wall. The field $s$ is an additional scalar coupled to
$\phi$ through a portal interaction. Motivated by generic renormalizable
scalar-portal extensions, we employ the finite-temperature potential
\cite{Quiros:1999jp,Profumo:2014opa,Tofighi:2015fia,Zhou:2020ojf}
\begin{align}
V(\phi,s,T)={}&
\frac{m_\phi^2(T)}{2}\phi^2
-\frac{a_\phi}{3}\phi^3
+\frac{\lambda_\phi}{4}\phi^4
\nonumber\\
&+\frac{m_s^2(T)}{2}s^2
+\frac{\lambda_s}{4}s^4
+\frac{\lambda_{\phi s}}{4}\phi^2s^2\,.
\label{eq:potential}
\end{align}
Here $m_\phi^2(T)$ and $m_s^2(T)$ include temperature-dependent corrections,
$\lambda_\phi$ and $\lambda_s$ are quartic self-couplings, $a_\phi$ is a
dimensionful cubic coefficient, and $\lambda_{\phi s}$ is the portal coupling.
The temperature is held at the nucleation value $T_n$ during the real-time
simulations.
Unless stated otherwise, all simulations use the benchmark parameters summarized in Table~\ref{tab:benchmark}.

For a fixed local value of $s$, the driving free-energy difference may be
written as
\begin{equation}
  \Delta V(s)=V(\phi_{\rm false},s,T_n)
             -V(\phi_{\rm true},s,T_n).
  \label{eq:local_free_energy_difference}
\end{equation}
Because of the $\phi^2s^2$ portal term, spatial variations in $s$ generate a
spatially varying $\Delta V$. Regions with a larger $|s|$ can reduce the local
driving pressure and may temporarily render the broken-phase minimum less
favorable. Regions with smaller $|s|$ exert a weaker obstruction and allow the
wall to reaccelerate. The even dependence on $s$ implies that both signs of the
additional field contribute to this effect.

\begin{figure}[t]
\centering
\safeincludegraphics[width=0.62\textwidth]{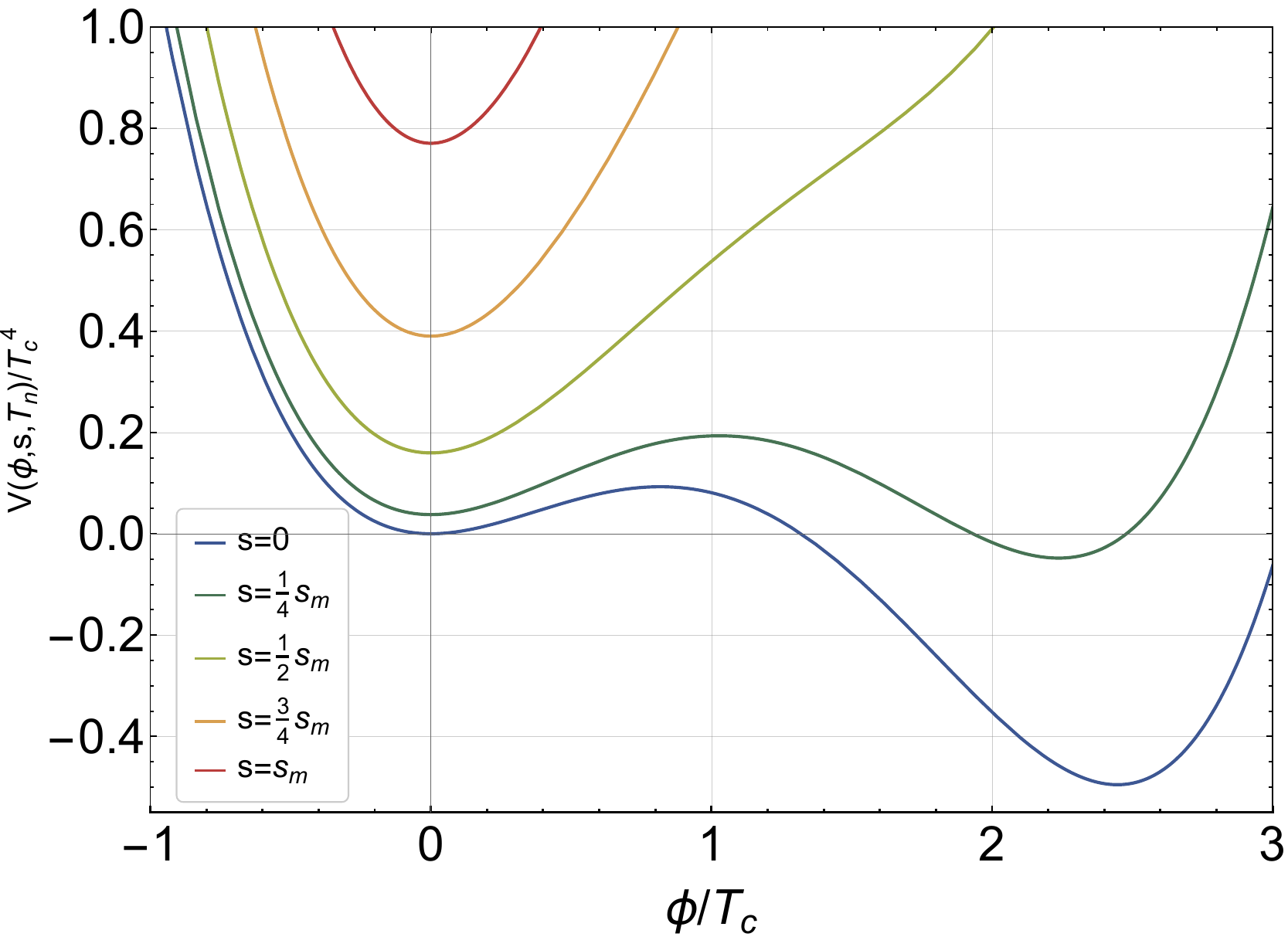}
\caption{Potential slices $V(\phi,s,T_n)$ as functions of $\phi/T_c$ for five
fixed values of $s$. Here $s_m$ denotes the maximum value attained by the
stochastic $s$ field in the representative realization used to define the
slices. The benchmark parameters are those in Table~\ref{tab:benchmark}, with
$\lambda_{\phi s}=1$. The variation of the relative vacuum depths illustrates
how the stochastic $s$ background makes the local driving pressure patch
dependent.}
\label{fig:potential}
\end{figure}

A related lattice study has considered a mass-acquiring additional scalar whose
suppression inside true-vacuum bubbles can release extra energy onto the wall
and strengthen the transition \cite{Li:2025zxa}. Here we emphasize the
complementary situation in which pre-existing spatial inhomogeneities of the
additional field intermittently suppress the local driving pressure and reduce
the wall propagation rate.

We focus on the post-nucleation propagation of the bubble wall rather than the
quantum nucleation process. The phase-transition field $\phi$ is therefore
initialized with a prescribed true-vacuum bubble, and its subsequent classical
evolution is followed on the lattice. In three spatial dimensions, the initial
profile is
\begin{align}
\phi(t_0,\mathbf{x})={}&\phi_{\rm false}
+\frac{\phi_{\rm true}-\phi_{\rm false}}{2}
\left[1-\tanh\!\left(\frac{r-R_0}{\ell_w}\right)\right],
\nonumber\\
\dot\phi(t_0,\mathbf{x})={}&0,
\qquad r=|\mathbf{x}-\mathbf{x}_0|.
\label{eq:phi_initial}
\end{align}
The radius $R_0$ is chosen above the expansion threshold, $\ell_w$ determines the initial wall thickness, and $t_0$ denotes the initial simulation time. Setting $\dot\phi=0$ avoids imposing an artificial
initial boost. The analogous $1+1$ dimensional setup uses two oppositely
oriented walls that delimit a true-vacuum interval.

The additional scalar field is initialized as
\begin{equation}
  s(0,\mathbf{x})
  =
  \bar{s}+\delta s(\mathbf{x}),
  \qquad
  \dot{s}(0,\mathbf{x})
  =
  \delta\dot{s}(\mathbf{x}),
  \label{eq:s_initial}
\end{equation}
where $\delta s$ and $\delta\dot{s}$ are sampled from a prescribed classical
stochastic ensemble. We set the homogeneous component $\bar s=0$ in all
simulations presented below.
\begin{table}[t]
\centering
\caption{Benchmark potential and bubble-profile parameters. The quantities
$R_0$ and $\ell_w$ are expressed in units of $T_c^{-1}$. The classical
fluctuation-amplitude parameter is fixed to $\As/T_c=0.8$ in the benchmark
runs.}
\label{tab:benchmark}
{
\begin{tabular}{cccccccc}
\toprule
  $m_\phi(T_n)/T_c$ & $\lambda_\phi$ & $m_s(T_n)/T_c$ &
$a_\phi/T_c$ & $\lambda_s$ & $T_n/T_c$ & $R_0T_c$ & $\ell_wT_c$ \\
\midrule
  $1$ & $0.5$ & $1$ & $1.632$ & $0.5$ & $0.8$ & $14.15$ & $1.71$ \\
\bottomrule
\end{tabular}}
\end{table}

The initial fluctuations are generated in Fourier space as Gaussian random
fields. For each independent Fourier mode, the field and velocity perturbations
are sampled subject to the reality conditions
\begin{equation}
  \delta s_{-\mathbf{k}}
  =
  \delta s_{\mathbf{k}}^{*},
  \qquad
  \delta\dot{s}_{-\mathbf{k}}
  =
  \delta\dot{s}_{\mathbf{k}}^{*}.
  \label{eq:stochastic_reality}
\end{equation}
Their variances are chosen as
\begin{align}
  \left\langle |\delta s_{\mathbf{k}}|^2 \right\rangle
  &=
  \frac{\As}{\omega_{\mathbf{k}}^2}
  \Theta\!\left(k_{\rm cut}-|\mathbf{k}|\right),
  &
  \left\langle |\delta\dot{s}_{\mathbf{k}}|^2 \right\rangle
  &=
  \As
  \Theta\!\left(k_{\rm cut}-|\mathbf{k}|\right),
  \label{eq:stochastic_variances}
\end{align}
where
\begin{equation}
  \omega_{\mathbf{k}}^2
  =
  \mathbf{k}^2+m_s^2(T_n).
  \label{eq:stochastic_frequency}
\end{equation}
 Gaussian classical-statistical ensembles specified through
equal-time two-point functions are widely used in real-time lattice simulations
\cite{Aarts:2001yn,Berges:2004yj,Berges:2013fga}. The parameter $\As$ controls the overall fluctuation
amplitude and is not interpreted as a thermodynamic temperature; we use
$\As/T_c=0.8$ in the benchmark simulations. The cutoff $k_{\rm cut}$ is chosen
below the lattice Nyquist momentum. In the resolution study, it is held fixed
in physical units so that lattice refinement does not alter the physical
ultraviolet content of the initial ensemble. The zero modes of the stochastic
perturbations are removed, so that
\begin{equation}
  \langle\delta s\rangle_{\mathbf{x}}
  =
  \langle\delta\dot{s}\rangle_{\mathbf{x}}
  =0.
  \label{eq:stochastic_zero_mode}
\end{equation}
Here $\langle\cdots\rangle_{\mathbf{x}}$ denotes a spatial average.

We do not derive the finite-cutoff spectrum used here from a specific
pre-transition history. It should instead be regarded as a controlled
parametrization of the stochastic state present at the onset of the
post-nucleation evolution. Such a state may arise, for example, from primordial
spectator fluctuations whose subsequent amplitude is determined by their
post-inflationary evolution, or from resonant amplification of an interacting
scalar field before the transition~\cite{LythWands2002,Greene1997,Khlebnikov1998,Antusch2023}.
The latter provides an explicit example in which amplified scalar fluctuations
remain dynamically important during bubble formation and expansion. Whether a
specific microscopic model produces the fluctuation amplitude and correlation
scales adopted here is model dependent.

The asymmetric treatment of the two scalar fields is primarily a controlled
modeling choice designed to isolate the influence of the inhomogeneous
$s$-field background on a bubble that has already nucleated. It should not be
interpreted as a statement that fluctuations of the phase-transition field
$\phi$ are strictly absent. Rather, the setup may be viewed as an approximation
to a class of histories for which
\begin{equation}
  \left\langle\delta s^2\right\rangle
  \gg
  \left\langle\delta\phi^2\right\rangle
  \label{eq:fluctuation_hierarchy}
\end{equation}
at the onset of the phase transition.
Such a hierarchy may result from different masses, couplings, or relaxation histories of the two fields before
the transition. Its microscopic realization is model dependent and lies
outside the scope of the present study.

\subsection{Lattice simulation setup}
\label{subsec:numerics}

The coupled classical equations are
\begin{align}
  \ddot\phi-\nabla^2\phi+\frac{\partial V}{\partial\phi}&=0,
  \label{eq:eom_phi}\\
  \ddot s-\nabla^2s+\frac{\partial V}{\partial s}&=0.
  \label{eq:eom_s}
\end{align}
We use periodic boundary conditions, second-order centered spatial differences,
and a staggered leapfrog integrator. The three-dimensional simulations use a cubic lattice with
$N^3=512^3$, $LT_c=200$, and $\Delta x\,T_c=0.39$. These simulations are
used to examine the resolved field profiles and possible lattice-aligned
distortions of the bubble shape.
The quantitative $1+1$ dimensional production runs
use $N=2048$ points with $LT_c=500$, corresponding to
$\Delta x\,T_c=500/2048\simeq0.244$, and
$\Delta t=0.025\,\Delta x$.
The smaller timestep used in the $1+1$-dimensional runs is a
conservative accuracy choice rather than a stability requirement; it suppresses
the temporal-integration error and energy drift, while the dependence on the
spatial discretization is examined separately through the lattice-resolution
study.

The leapfrog update and the construction of the first half step are summarized
in appendix~\ref{app:algorithm}. The analysis interval is always restricted to
times before signals can interact with their nearest periodic images.

A wall position $x_w(t)$ is defined by the midpoint contour
\begin{equation}
  \phi\bigl(t,x_w(t)\bigr)
  =\phi_*\equiv\frac{\phi_{\rm false}+\phi_{\rm true}}{2}.
  \label{eq:wall_midpoint}
\end{equation}
For the one-dimensional bubble, the left and right contours, $x_L(t)$ and
$x_R(t)$, define the effective half-size
\begin{equation}
  R(t)=\frac{x_R(t)-x_L(t)}{2}.
  \label{eq:bubble_half_size}
\end{equation}
The primary propagation observable is defined operationally as the slope of a
linear least-squares fit over a stated time interval,
\begin{equation}
  R(t)\simeq R_{\rm fit}+\vprop\,t,
  \qquad t_1\leq t\leq t_2.
  \label{eq:vprop}
\end{equation}
For an exactly linear trajectory, this agrees with the endpoint average
$[R(t_2)-R(t_1)]/(t_2-t_1)$; the two quantities need not coincide for the
intermittent trajectories encountered at strong coupling. The fit quality is
reported together with the slope. We use the term terminal velocity only when
the wall trajectory is demonstrably stationary. At strong coupling, $\vprop$
instead summarizes intermittent motion that can include slowing,
reacceleration, and temporary blocking.

In one spatial dimension, the total energy density is
\begin{equation}
\begin{aligned}
\rho_{\rm tot}(t,x)={}&
\frac12\dot\phi^2+\frac12(\partial_x\phi)^2
+\frac12\dot s^2+\frac12(\partial_xs)^2
+V(\phi,s,T_n).
\end{aligned}
\label{eq:total_energy_density}
\end{equation}
To track wave-like excitations of the additional field, we define
\begin{equation}
  \rhodyn(t,x)=\frac12\dot s^2+\frac12(\partial_xs)^2.
  \label{eq:s_dynamic_energy}
\end{equation}
This quantity excludes the local potential and portal energy, whose assignment
to either scalar sector is not unique, and isolates the kinetic and gradient
energy carried by excited $s$ modes.

For each wall, we introduce the outward-oriented coordinate
\begin{equation}
  \xi=n_w\,[x-x_w(t)],
  \label{eq:wall_frame_coordinate}
\end{equation}
so that $\xi>0$ denotes the false-vacuum side in front of the wall, whereas
$\xi<0$ denotes the bubble interior behind it.

The stochastic $s$ field already carries a nonzero kinetic and gradient energy
away from the wall. Let $d_w(x,t)$ denote the periodic distance from $x$ to the
nearest of the two walls, and define the far-field sampling set
\begin{equation}
  \mathcal{B}(t)
  =
  \bigl\{x\,\big|\,\xi_1\leq d_w(x,t)\leq\xi_2\bigr\}.
  \label{eq:background_sampling_set}
\end{equation}
For each saved time and each stochastic realization, we estimate a single
instantaneous far-field background level from the lattice sites in
$\mathcal{B}(t)$. Specifically, we collect the values of
$\rho_s^{\rm dyn}(t,x_j)$ at all lattice sites $x_j\in\mathcal{B}(t)$,
order these values, and take their median, namely the middle value of the
ordered sample:
\begin{equation}
  \rho_{\rm bg}^{\rm dyn}(t)
  =
  \operatorname{median}
  \left\{
    \rho_s^{\rm dyn}(t,x_j)
    \,\middle|\,
    x_j\in\mathcal{B}(t)
  \right\}.
  \label{eq:rho_bg_dynamic}
\end{equation}
The median is used instead of the spatial mean because it is less sensitive
to isolated peaks or residual wave packets in the far-field region. The same
background value $\rho_{\rm bg}^{\rm dyn}(t)$ is subtracted from the front- and
behind-the-wall profiles for that time and realization, so that both sides are
compared relative to a common reference level.

\begingroup
We then integrate the positive excess over two wall-frame intervals of equal
physical width. The parameter $\delta$ denotes the integration half-width
measured outward from the wall: $0\leq\xi\leq\delta$ covers the false-vacuum
side in front of the wall, while $-\delta\leq\xi\leq0$ covers the bubble
interior behind it. We choose
\begin{equation}
  \delta = 10\,T_c^{-1}.
  \label{eq:integration_half_width}
\end{equation}
We define the positive excess as
\begin{equation}
  \delta\rho_s^{+}(t,\xi)
  \equiv
  \bigl[\rho_s^{\rm dyn}(t,\xi)-\rho_{\rm bg}^{\rm dyn}(t)\bigr]_+
  =
  \max\!\left(
  \rho_s^{\rm dyn}(t,\xi)-\rho_{\rm bg}^{\rm dyn}(t),\,0
  \right).
  \label{eq:positive_excess}
\end{equation}
The integrated positive excesses are then
\begin{align}
  I_{\rm front}(t)
  &=
  \int_0^\delta
  \delta\rho_s^{+}(t,\xi)
  \,\dd\xi,
  \label{eq:I_front}
  \\
  I_{\rm behind}(t)
  &=
  \int_{-\delta}^{0}
  \delta\rho_s^{+}(t,\xi)
  \,\dd\xi.
  \label{eq:I_behind}
\end{align}
Thus, only energy-density values above the estimated far-field background
contribute to $I_{\rm front}$ and $I_{\rm behind}$. The positive-part
prescription is used because the background-subtracted quantity can fluctuate
both above and below the estimated far-field level. Negative values do not
represent negative energy; they simply correspond to local values below the
estimated background. Retaining only the positive part avoids cancellations
between localized excess-energy peaks and sub-background troughs, so that
$I_{\rm front}$ and $I_{\rm behind}$ measure the wall-associated dynamical-energy
excess on the two sides of the interface.

For each saved time, the outward-oriented profiles of the two walls are
smoothed with a five-point moving average and then averaged at fixed $\xi$
before the integrated quantities are evaluated. The corresponding
front--behind asymmetry is
\begin{equation}
  \AFB(t)
  =
  \frac{I_{\rm front}(t)-I_{\rm behind}(t)}
       {I_{\rm front}(t)+I_{\rm behind}(t)},
  \label{eq:AFB}
\end{equation}
whenever $I_{\rm front}(t)+I_{\rm behind}(t)>0$. By construction,
$-1\leq\AFB\leq1$. Positive values indicate that the excess dynamical energy
is preferentially located on the false-vacuum side in front of the wall,
whereas negative values indicate a behind-the-wall bias.
\endgroup

The analysis choices entering this construction are intended only to define a simple quantitative measure of the front--behind energy redistribution and are not optimized to maximize its correlation with the wall propagation rate. We have verified that the qualitative behavior of $\AFB$ is unchanged under reasonable variations of the integration half-width, background prescription, smoothing window, time-averaging interval, wall-location criterion, and spectator-energy-excess definition, as well as under lattice refinement.

Quantitative observables are obtained from ensembles of independent stochastic
realizations generated with different random seeds. The ensemble size depends
on the scan and is specified together with the corresponding result. Within
each scan, the same seed list is reused across the parameter values being
compared whenever applicable, thereby reducing sampling noise in direct
parameter-to-parameter comparisons. Ensemble results are reported as sample
means, with error bars denoting the standard error of the mean unless stated
otherwise.

\section{Bubble-wall dynamics}
\label{sec:dynamics}

The three-dimensional simulations serve as a geometry test. They determine
whether the additional-field backreaction survives for a spherical bubble and
a spatially inhomogeneous environment, rather than being a consequence of the
one-dimensional reduction.

\subsection{Three-dimensional bubble evolution}
\label{subsec:3d_evolution}

When the additional scalar field is coupled to the
phase-transition field, the portal contribution
\begin{equation}
  V_{\rm int}(\phi,s)
  =
  \frac{\lambda_{\phi s}}{4}\phi^2s^2,
\end{equation}
makes the local free-energy difference across the wall dependent on the
inhomogeneous $s$ background. Schematically, the local driving pressure can
be written as
\begin{equation}
  \Delta p_{\rm eff}(s)
  =
  V(\phi_{\rm false},s)-V(\phi_{\rm true},s)
  \simeq
  \Delta p_0
  -
  \frac{\lambda_{\phi s}}{4}
  \left(\phi_{\rm true}^2-\phi_{\rm false}^2\right)s^2,
  \label{eq:local_driving_pressure}
\end{equation}
where $\Delta p_0$ denotes the driving pressure in the absence of the
additional-field background. For positive $\lambda_{\phi s}$, regions with
large $|s|$ therefore reduce the local preference for the true vacuum,
irrespective of the sign of $s$. Since the stochastic field varies in space,
different portions of the spherical wall encounter different effective
driving pressures. The portal-induced backreaction is consequently
inhomogeneous and time dependent, leading to local distortions of the wall
and realization-dependent changes in its propagation rate.

\begin{figure}[t]
\centering
\safeincludegraphics[width=\textwidth]{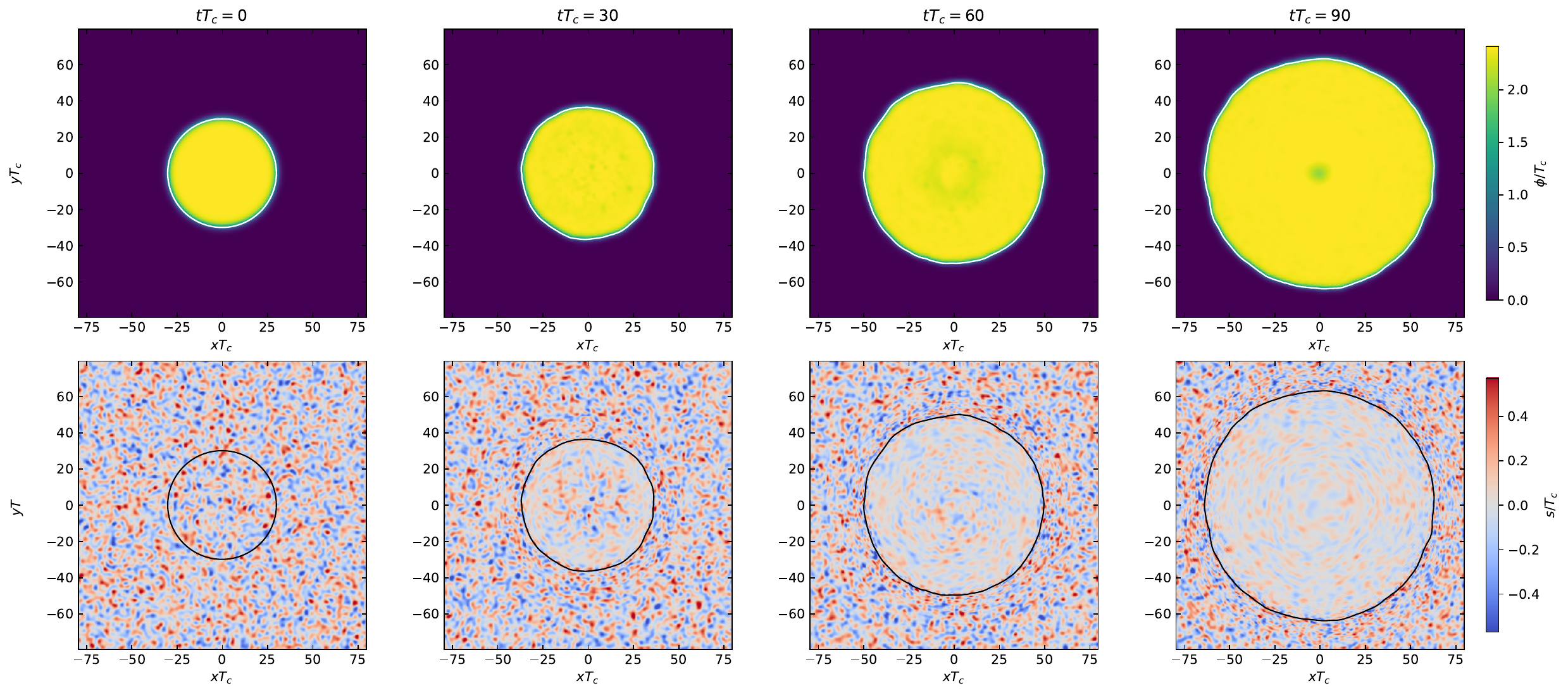}
\caption{Representative two-dimensional slices through the center of the three-dimensional bubble at $tT_c=0$, $30$, $60$, and $90$. The upper panels show the phase-transition field $\phi/T_c$, while the lower panels show the spectator field $s/T_c$. The contour denotes the bubble wall defined by $\phi=\phi_{*}$.}
\label{fig:3d_profiles}
\end{figure}

The spatially varying driving pressure also produces a nonuniform evolution of the bubble surface. Figure~\ref{fig:3d_profiles} shows representative two-dimensional slices through the bubble center at several stages of the evolution. The upper panels display the phase-transition field $\phi$, while the lower panels show the corresponding spectator-field configuration $s$. The contour $\phi=\phi_{*}$ indicates the instantaneous position of the bubble wall.

The bubble remains approximately spherical on large scales, while local deformations gradually develop as the wall propagates through the inhomogeneous spectator-field background. Different portions of the wall encounter regions with different values of $s^2$ and therefore experience different effective driving pressures. Consequently, some wall segments are locally delayed relative to others, producing the irregular structure visible along the bubble surface at later times. The evolution of the wall morphology thus provides a direct geometric manifestation of the spatially inhomogeneous portal-induced backreaction.

The slower expansion observed in the coupled simulation should thus not be
interpreted as the action of a spatially uniform viscous force. At the
microscopic level, the coupled two-field system evolves conservatively:
energy released by the advancing wall can be transferred into kinetic and
gradient excitations of the $s$ field, while the portal interaction also
allows energy to be exchanged back and forth between the two scalar sectors.
The oscillatory structures behind the wall are a manifestation of this
coherent field dynamics. Only after coarse graining over the rapidly varying
field excitations does the cumulative loss of energy from the propagating wall
appear as an effective friction-like backreaction.

A transient shrink--reexpand episode may occur when the initially prescribed
bubble overlaps a region in which $s^2$ is sufficiently large that the local
vacuum driving pressure is strongly reduced. During this stage, the outward
vacuum pressure may temporarily fail to overcome the inward effects associated
with the wall tension, the initial curvature, and the relaxation of the
non-equilibrium wall profile. As the $s$ configuration evolves and the wall
moves away from the strongly suppressing region, the effective driving
pressure can recover, allowing the bubble to resume its expansion. The
transient therefore reflects the local and dynamical nature of the portal
backreaction rather than a permanent reversal of the phase transition.

\begin{figure}[t]
\centering
\safeincludegraphics[width=0.70\textwidth]{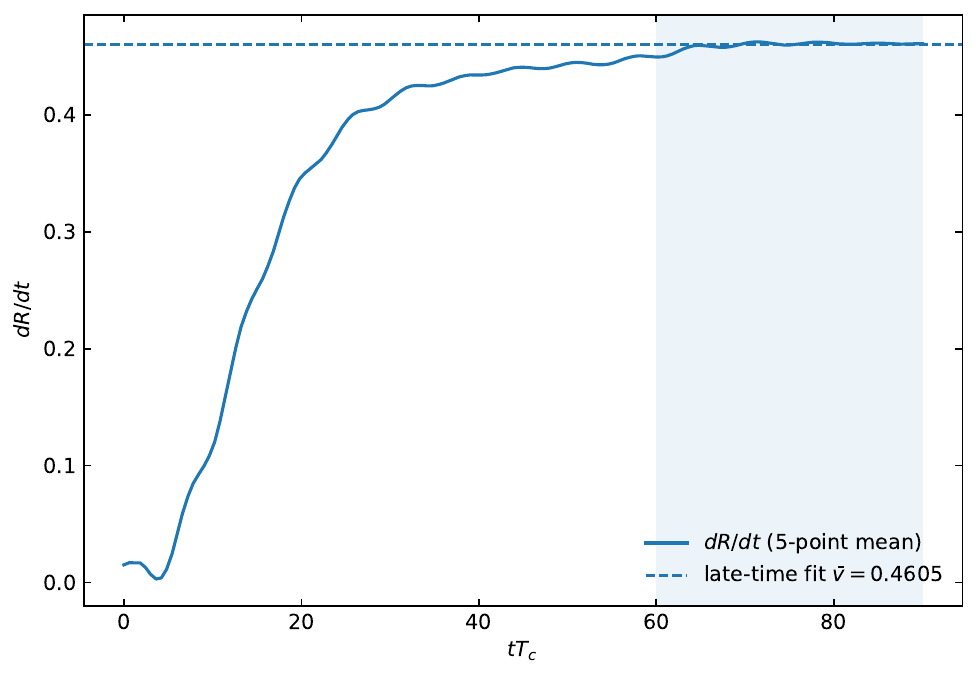}
\caption{Evolution of the three-dimensional bubble propagation rate
extracted from the volume-equivalent radius $R(t)$ for the representative
coupled run with $\lambda_{\phi s}=2$. The solid curve shows $dR/dt$ after a
five-point moving average. The dashed horizontal line denotes the finite-time
propagation rate $\bar v_{\rm prop}=0.46$ obtained from a linear fit to
$R(t)$ over $60\leq tT_c\leq90$, and the shaded region indicates the fitting
interval.}
\label{fig:three_dimensional_velocity}
\end{figure}

Figure~\ref{fig:three_dimensional_velocity} shows the corresponding
global propagation-rate history extracted from the volume-equivalent bubble
radius $R(t)$. Using a volume-based radius avoids selecting a particular
Cartesian direction and therefore reduces the sensitivity of the velocity
measurement to local distortions of the wall. After the initial relaxation,
the propagation rate approaches an approximately stationary value over the
interval $60\leq tT_c\leq90$. A linear fit to $R(t)$ in this interval gives
$\bar v_{\rm prop}=0.46$. This global measure is used here as the
representative three-dimensional propagation observable.

The three-dimensional simulations establish that the portal-induced
suppression and intermittency persist for a curved, expanding bubble in an
inhomogeneous background. They also reveal geometric effects that are absent
in one dimension, including angular variations of the local driving pressure
and departures from exact spherical symmetry. However, a single
three-dimensional realization is not sufficient to determine the statistical
dependence on $\lambda_{\phi s}$. The quantitative coupling scan and ensemble
averages are therefore obtained from the one-dimensional simulations discussed
below.

\subsection{One-dimensional wall propagation}
\label{subsec:1d_dynamics}

The one-dimensional geometry resolves the wall-normal dynamics at substantially
lower computational cost. It permits longer evolution times, a denser portal-coupling scan, multiple stochastic realizations, and systematic resolution
tests.

\begin{figure}[t]
\centering
\safeincludegraphics[width=0.31\textwidth]{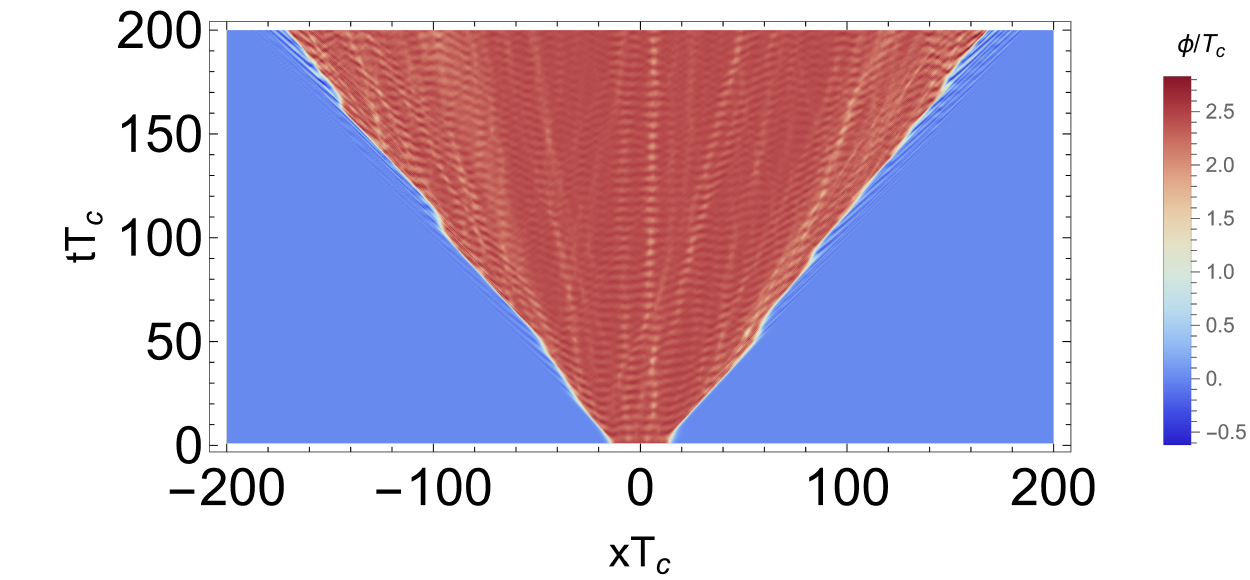}\hfill
\safeincludegraphics[width=0.31\textwidth]{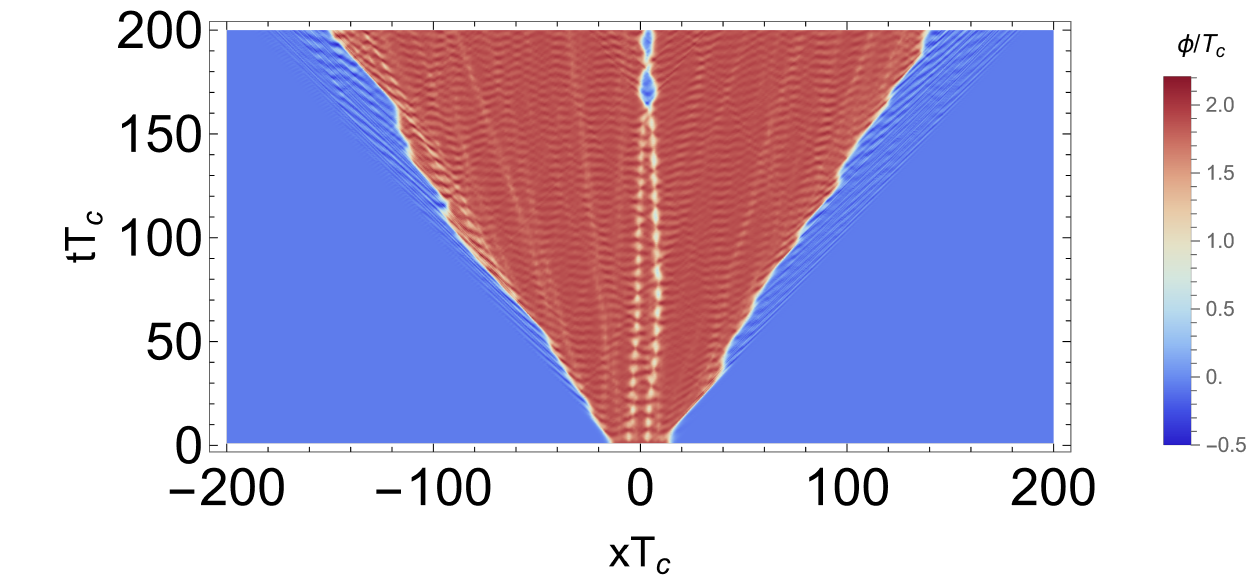}\hfill
\safeincludegraphics[width=0.31\textwidth]{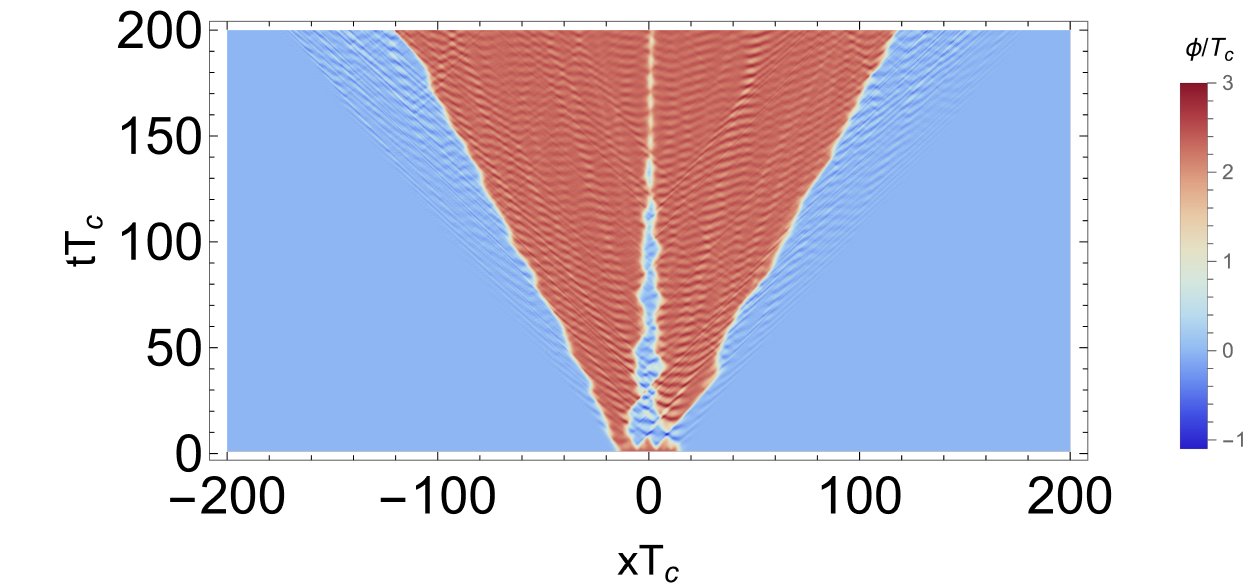}
\caption{Representative $1+1$ dimensional evolutions for
$\lambda_{\phi s}=0.65$, $1.1$, and $1.8$ from left to right. The runs use
$N=2048$, $LT_c=500$, $\Delta x\,T_c\simeq0.244$, and
$\Delta t=0.025\,\Delta x$. These individual realizations illustrate the
changing wall trajectories; quantitative statements must be based on the
multi-seed scan in figure~\ref{fig:ensemble_velocity_scan}.}
\label{fig:one_dimensional_spacetime}
\end{figure}

Figure~\ref{fig:one_dimensional_spacetime} shows representative wall
trajectories in the one-dimensional simulations. At weaker coupling, the two
outgoing walls follow nearly straight trajectories after the initial transient,
indicating an approximately steady propagation rate over the displayed interval.
At intermediate coupling, the mean slope is reduced and local changes of slope
appear as the walls encounter different regions of the inhomogeneous $s$-field
background. At stronger coupling, the motion becomes more intermittent, with
pronounced deviations from linear propagation, including temporary blocking
followed by reacceleration. These single realizations provide a qualitative view
of the coupling-dependent wall dynamics and motivate the finite-time observable
introduced in eq.~\eqref{eq:vprop}; the corresponding statistical dependence on
$\lambda_{\phi s}$ is established by the multi-realization scan below.

We have additionally checked the late-time behavior in a larger
$1+1$-dimensional volume with $N=3072$ and $LT_c=750$, keeping the same spatial
resolution, $\Delta x\,T_c=0.244$. Extending the evolution to
$tT_c=300$ confirms that the weak-coupling trajectories approach an approximately
stationary propagation regime, whereas the stronger-coupling trajectories continue
to exhibit intermittent variations in their propagation rate.

\begin{figure}[t]
  \centering
  \safeincludegraphics[width=\textwidth]
  {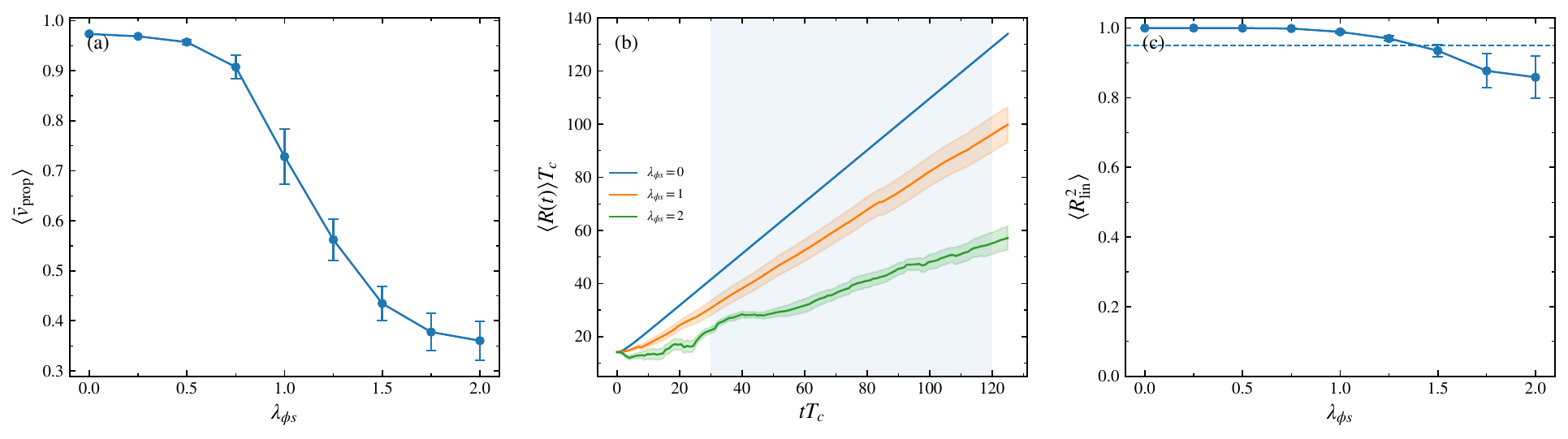}
  \caption{
  Dependence of the bubble-wall propagation on the portal coupling.
  Left: ensemble-averaged finite-time propagation rate.
  Center: ensemble-averaged radius histories for representative coupling
  strengths.
  Right: linear-fit quality $R_{\rm lin}^{2}$.
  Each point is averaged over ten stochastic realizations, and the error bars
  and shaded bands denote the standard error of the mean.
  }
  \label{fig:ensemble_velocity_scan}
\end{figure}

Figure~\ref{fig:ensemble_velocity_scan} presents the main quantitative test of
the portal-induced suppression of bubble-wall propagation. For each value of
$\lambda_{\phi s}$, we perform ten simulations with independent stochastic
realizations of the initial $s$-field background. The same stochastic-spectrum
parameters,
$\As/T_c=0.8$ and $k_{\rm cut}=0.25\,k_{\rm Ny}$, are used throughout the
scan.

For each realization, the finite-time propagation rate is extracted from a
linear fit to the midpoint-contour radius,
\begin{equation}
  R(t)
  \simeq
  R_{\rm fit}
  +
  \bar v_{\rm prop}\,t,
  \qquad
  30\leq tT_c\leq120.
  \label{eq:radius_linear_fit}
\end{equation}
The left panel shows that the ensemble-averaged propagation rate decreases
systematically as the portal coupling is increased. The suppression is weak
at small $\lambda_{\phi s}$ but becomes pronounced once the portal interaction
substantially modifies the local driving pressure. The statistical uncertainty
also grows at stronger coupling, reflecting the increasing sensitivity of the
wall motion to the particular realization of the inhomogeneous $s$ background.

The central panel shows representative ensemble-averaged radius histories.
In the decoupled case, the radius grows nearly linearly throughout the fitting
interval. Increasing $\lambda_{\phi s}$ reduces the mean slope and produces
more pronounced deviations from uniform propagation. The shaded regions
denote the standard error of the ensemble mean and therefore quantify the
uncertainty in the estimated mean trajectory; they do not represent the full
spread of individual realizations.

To assess whether a single fitted slope provides an adequate description of
the wall motion, we use the coefficient of determination
\begin{equation}
  R_{\rm lin}^{2}
  =
  1-
  \frac{
    \sum_i
    \left[
      R(t_i)-R_{\rm fit}(t_i)
    \right]^2
  }{
    \sum_i
    \left[
      R(t_i)-\overline{R}
    \right]^2
  },
  \label{eq:Rlin_definition}
\end{equation}
where
\begin{equation}
  R_{\rm fit}(t_i)
  =
  R_{\rm fit}
  +
  \bar v_{\rm prop}\,t_i
\end{equation}
is the fitted linear trajectory and $\overline{R}$ is the mean radius within
the fitting interval.

Values of $R_{\rm lin}^{2}$ close to unity indicate that $R(t)$ is well
approximated by a linear function over the selected time interval. A decreasing
$R_{\rm lin}^{2}$ signals stronger curvature or intermittency in the radius
history, such as alternating periods of suppression and renewed expansion.
Therefore, $R_{\rm lin}^{2}$ measures the quality of the finite-time linear
description; it does not by itself demonstrate that the wall has reached a
universal terminal velocity.

The right panel shows that the linear approximation remains highly accurate at
weak coupling, whereas its quality gradually deteriorates in the stronger
coupling regime. We therefore use $\bar v_{\rm prop}$ as a finite-time summary
of the wall motion, while recognizing that a single constant velocity becomes
a less complete description when the portal-induced blocking is strongly
intermittent.

To isolate the dependence on the strength of the stochastic background, we fix
the portal coupling at $\lambda_{\phi s}=1$ and rescale the initial
fluctuations according to
\begin{equation}
  \delta s(\mathbf{x})
  =
  \alpha_s\,\delta s_0(\mathbf{x}),
  \qquad
  \delta\dot{s}(\mathbf{x})
  =
  \alpha_s\,\delta\dot{s}_0(\mathbf{x}).
  \label{eq:stochastic_amplitude_rescaling}
\end{equation}
The spectral shape, reference amplitude, and physical momentum cutoff are kept
fixed. For the classical spectrum adopted here, this rescaling multiplies both
field and momentum variances by $\alpha_s^2$ and is therefore equivalent to
replacing the amplitude parameter by
$\As^{\rm eff}=\alpha_s^2\As$, without changing the spectral shape.

We scan $41$ equally spaced amplitudes over
$0\leq\alpha_s\leq2$, with a spacing $\Delta\alpha_s=0.05$. For each value of
$\alpha_s$, the result is averaged over ten independent stochastic
realizations. The finite-time propagation rate is extracted from the
midpoint-contour radius over the same fitting interval,
$30\leq tT_c\leq120$, used in the portal-coupling scan.

\begin{figure}[t]
  \centering
  \safeincludegraphics[width=0.62\textwidth]
  {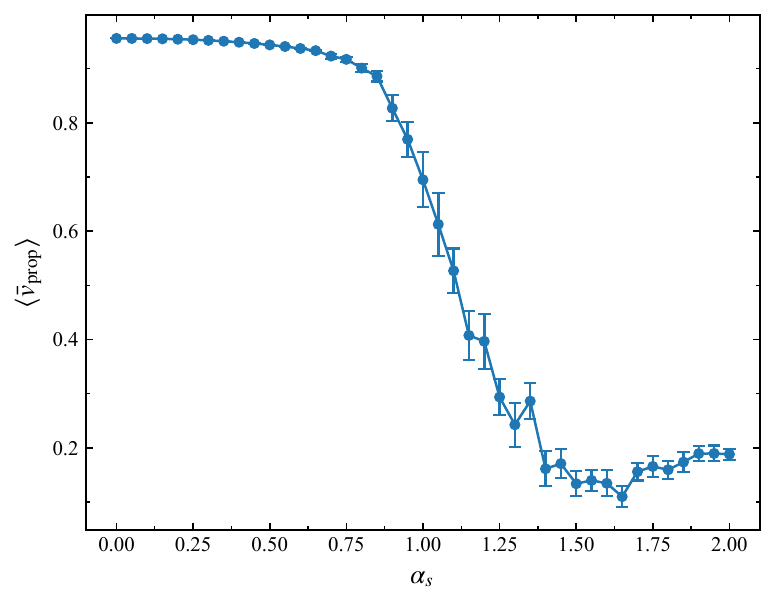}
  \caption{
  Ensemble-averaged finite-time propagation rate as a function of the
  stochastic-background amplitude $\alpha_s$ at fixed
  $\lambda_{\phi s}=1$. Each point is averaged over ten independent
  realizations, and the error bars denote the standard error of the mean.
  }
  \label{fig:fluctuation_amplitude_scan}
\end{figure}

Figure~\ref{fig:fluctuation_amplitude_scan} shows that the wall propagation is
strongly sensitive to the amplitude of the stochastic background. In the weak
background regime, the suppression is gradual. The averaged propagation rate
decreases only mildly from
$\langle\bar v_{\rm prop}\rangle\simeq0.956$ at $\alpha_s=0$ to approximately
$0.933$ at $\alpha_s=0.65$. The small error bars in this region show that
different realizations produce very similar wall trajectories.

A much stronger suppression appears for
$0.7\lesssim\alpha_s\lesssim1.3$. Across this interval,
$\langle\bar v_{\rm prop}\rangle$ decreases from approximately $0.923$ at
$\alpha_s=0.7$ to $0.695$ at $\alpha_s=1$, and further to approximately
$0.243$ at $\alpha_s=1.3$. The simultaneous broadening of the error bars
indicates that the wall motion also becomes increasingly sensitive to the
spatial realization of the stochastic background. This behavior is consistent
with a crossover from weak perturbations of the driving pressure to a regime
in which localized regions of large $s^2$ can substantially delay the wall.

For $\alpha_s\gtrsim1.4$, the fitted propagation rate remains within a
low-velocity range of approximately $0.1$--$0.2$, but its dependence on
$\alpha_s$ is no longer strictly monotonic. For example, the mean rate reaches
approximately $0.134$ at $\alpha_s=1.5$, decreases to approximately $0.110$ at
$\alpha_s=1.65$, and subsequently rises to approximately $0.188$ at
$\alpha_s=2$.

This behavior reflects the transition from a perturbative suppression regime
to an intermittently blocked regime. At the initial time,
$s(0,\mathbf{x})=\alpha_s s_0(\mathbf{x})$, so the portal-induced correction
to the local driving pressure initially scales as
$\alpha_s^2s_0^2(\mathbf{x})$. The subsequent nonlinear evolution does not
preserve this scaling exactly, but the initial relation explains why increasing
$\alpha_s$ rapidly raises the probability of encountering strongly suppressing
patches. At sufficiently large $\alpha_s$, some spatial patches may reduce the
local driving pressure to nearly zero or may even temporarily reverse its
sign. The wall then alternates between periods of propagation, local blocking,
and renewed advance as the inhomogeneous $s$ configuration evolves.

In this regime, the fitted value of $\bar v_{\rm prop}$ is controlled not only
by the instantaneous strength of the obstruction, but also by the number,
duration, and timing of the blocking episodes within the finite fitting
interval. Consequently, a larger $\alpha_s$ need not produce a smaller fitted
slope for every point in the scan. The modest rise observed at the largest
amplitudes should therefore not be interpreted as a systematic recovery of
the wall speed. Consistently, the linear-fit quality decreases to
$\langle R_{\rm lin}^2\rangle\simeq0.5$--$0.8$, showing that the trajectories
are strongly intermittent and are no longer accurately characterized by a
single constant propagation rate.

The amplitude scan therefore shows that the initial stochastic-background
strength controls the onset of the portal-induced backreaction. The response
is approximately smooth in the weak-background regime, becomes rapidly
stronger when local regions begin to approach the blocking threshold around
$\alpha_s\sim1$, and enters a nonlinear strong-blocking regime at larger
amplitudes. In the latter regime, $\bar v_{\rm prop}$ should be regarded as a
finite-time summary of the net wall displacement rather than as a
well-defined terminal wall velocity.
\section{Dynamical-energy redistribution and wall-frame structure}
\label{sec:energy}

The propagation rate provides a kinematic characterization of the bubble wall,
but it does not by itself identify the microscopic origin of the slowdown. In
the present conservative two-field system, a reduction of the wall propagation
rate must be accompanied by a redistribution of energy among the
phase-transition field, the portal interaction, and the kinetic and gradient
excitations of the additional scalar field. Determining where these excitations
are deposited relative to the moving interface is therefore essential for
understanding how the inhomogeneous $s$ background backreacts on the wall.

This redistribution is also relevant to the phenomenology of cosmological
first-order phase transitions. In three spatial dimensions, inhomogeneous
scalar-field gradients and kinetic stresses contribute to the anisotropic stress
that sources gravitational waves. Consequently, the amount, spatial
localization, and persistence of the excited scalar-field energy may affect the
energy budget and coherence of the gravitational-wave source during bubble
growth and collision
\cite{Espinosa:2010hh,Hindmarsh:2013xza,Caprini:2019egz,
Wang:2020jrd}. The present one-dimensional simulations do not determine a
gravitational-wave spectrum; instead, they provide a microscopic wall-frame
diagnostic of the energy redistribution that would enter a future
three-dimensional calculation. In particular, energy deposited behind the wall
forms an excitation wake, whereas accumulation on the false-vacuum side
modifies the region into which the wall is attempting to propagate.

\begin{figure}[t]
\centering
\safeincludegraphics[width=0.325\textwidth]{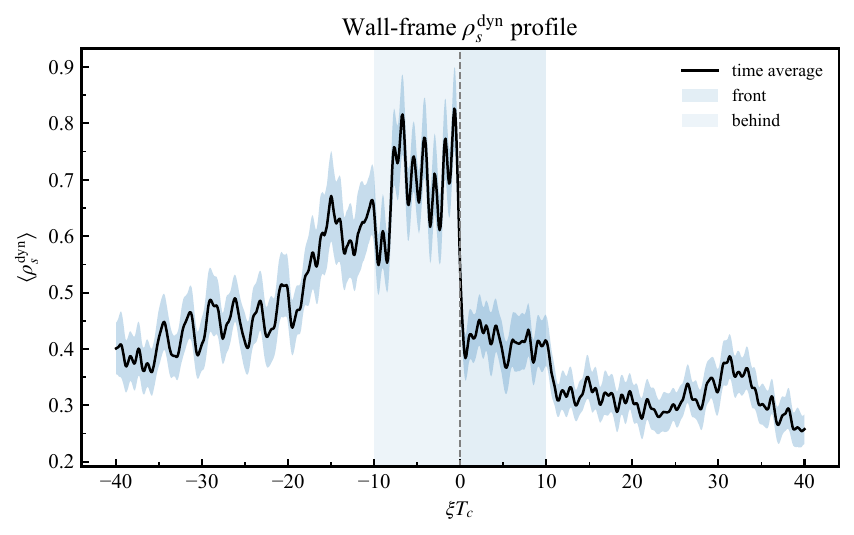}\hfill
\safeincludegraphics[width=0.325\textwidth]{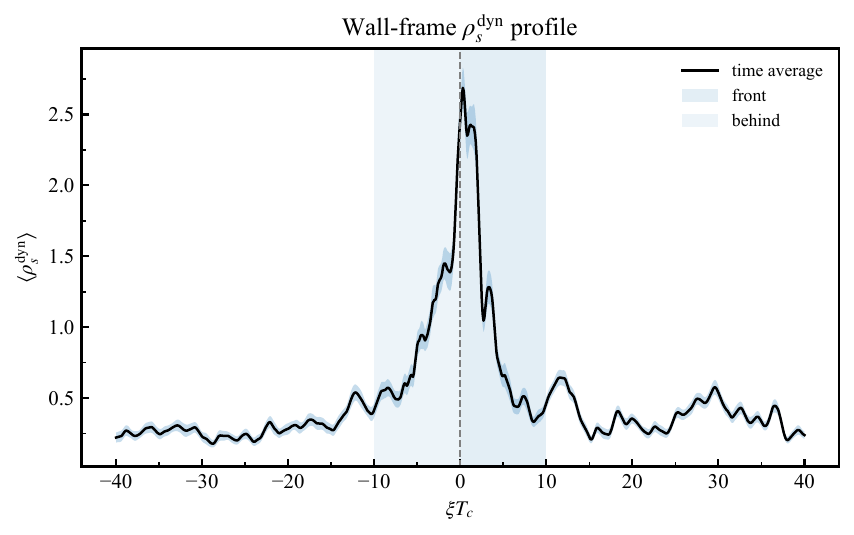}\hfill
\safeincludegraphics[width=0.325\textwidth]{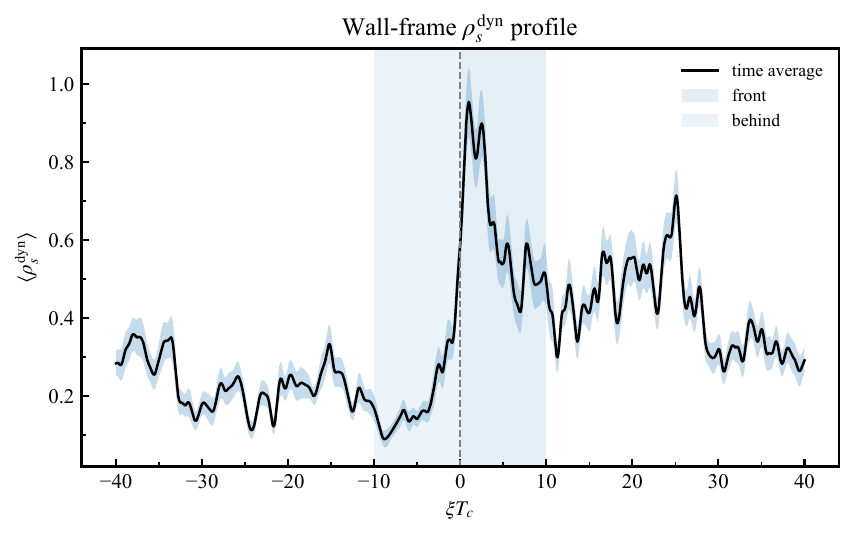}
\caption{Wall-frame profiles of the dynamical spectator-field
energy density $\rhodyn$ for $\lambda_{\phi s}=0.5$, $1$, and $1.5$, from
left to right. Light and dark shading indicate the regions behind and in front
of the wall, respectively. All panels use the same vertical scale.}
\label{fig:rhos_wallframe_profiles}
\end{figure}

The spatially resolved three-dimensional configurations in
Fig.~\ref{fig:3d_profiles} already illustrate that the spectator-field
structure around the propagating wall is highly nonuniform and evolves with
time. To display the corresponding wall-normal energy distribution more
directly, Fig.~\ref{fig:rhos_wallframe_profiles} shows representative
wall-frame profiles of the dynamical spectator-field energy density
$\rhodyn$ for weak, intermediate, and strong portal couplings. At weak
coupling the excess dynamical energy is concentrated predominantly behind the
wall. Around the intermediate-coupling crossover the distribution moves closer
to the wall and becomes more realization dependent, whereas at stronger
coupling a pronounced excess develops on the false-vacuum side.

Together, Figs.~\ref{fig:3d_profiles} and
\ref{fig:rhos_wallframe_profiles} provide complementary views of the same
inhomogeneous backreaction: the former shows the spatially resolved field
configuration around the three-dimensional bubble, while the latter displays
the corresponding wall-normal redistribution of dynamical spectator-field
energy. For quantitative comparisons across portal couplings and stochastic
realizations, we therefore use $\AFB$ below as a simple integrated measure of
this front--behind redistribution rather than as a substitute for the full
profiles.

\begin{figure}[t]
\centering
\safeincludegraphics[width=0.95\textwidth]{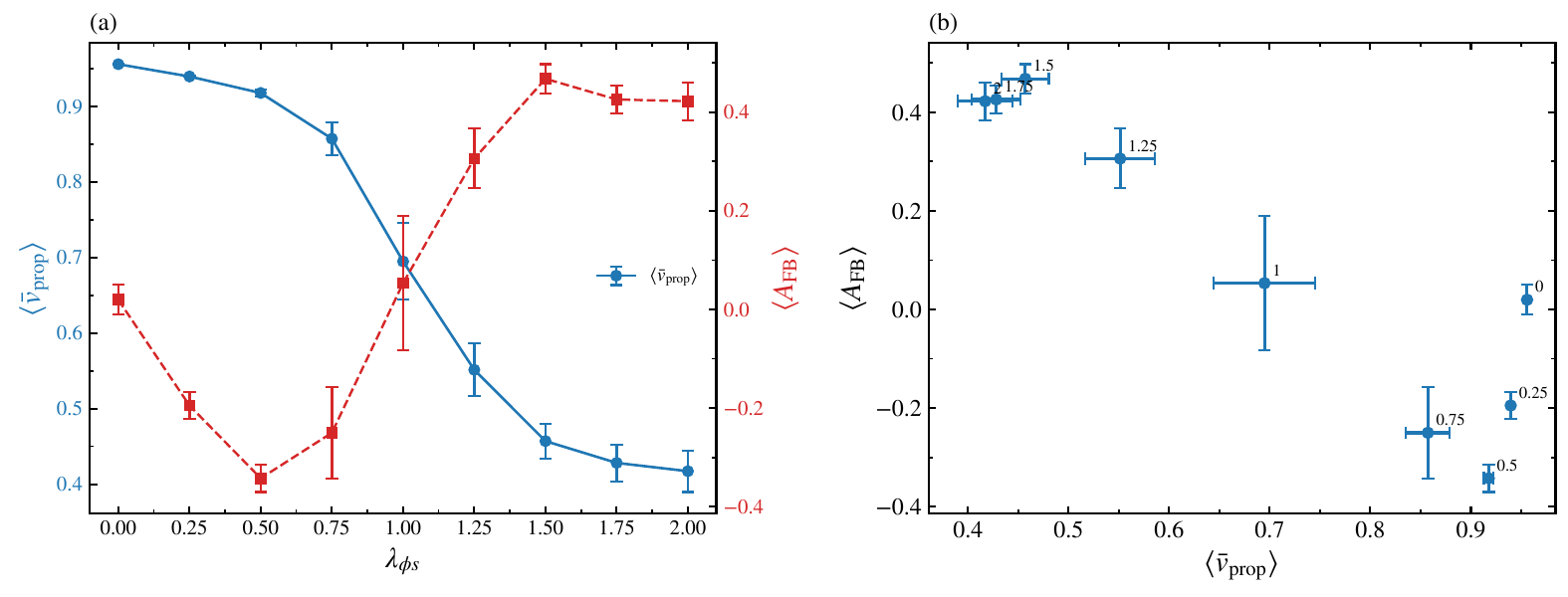}
\caption{Ensemble-averaged relation between bubble-wall propagation and the
front--behind asymmetry of the dynamical $s$-field energy.
Left: $\vev{\vprop}$ and $\vev{\AFB}$ as functions of
$\lambda_{\phi s}$.
Right: $\vev{\AFB}$ versus $\vev{\vprop}$, with the portal coupling indicated
beside each point.
Error bars denote the standard error of the mean over ten stochastic
realizations.}
\label{fig:AFB_summary}
\end{figure}
\begingroup
To quantify the redistribution visible in
Fig.~\ref{fig:rhos_wallframe_profiles}, we use the local observable $\rhodyn$
defined in eq.~\eqref{eq:s_dynamic_energy} and the front--behind asymmetry
$\AFB$ defined in eq.~\eqref{eq:AFB}. The latter compresses the wall-frame
energy distribution into a single ensemble-level measure and is used only to
facilitate systematic comparisons across couplings and realizations.

For the results shown in Fig.~\ref{fig:AFB_summary}, the baseline analysis uses
the integration half-width $\delta T_c=10$. The far-field background is
estimated from lattice points satisfying $40\leq d_wT_c\leq100$ using the
median of $\rhodyn$. The same instantaneous background $\rhobg(t)$ is
subtracted from the profiles on both sides of the wall before retaining the
positive excess $\delta\rho_s^{+}(t,\xi)$. The asymmetry is time-averaged over
$30\leq tT_c\leq120$ for each realization and subsequently averaged over ten
independent stochastic realizations. The same prescription is applied to all
values of $\lambda_{\phi s}$.
\endgroup

We have also checked the robustness of this diagnostic by varying the integration half-width, the background region and estimator, the smoothing window, the averaging interval, the wall-location criterion, and the definition of the spectator-energy excess. These variations can moderately change the numerical magnitude of $\AFB$, but do not alter the qualitative coupling-dependent front--behind redistribution discussed below.

The longer-time simulations discussed in
Sec.~\ref{subsec:1d_dynamics} further show that intermittent wall dynamics
persist at stronger coupling. We therefore do not interpret the profiles in
Fig.~\ref{fig:rhos_wallframe_profiles} as evidence for a universal
time-independent shell shape; rather, they characterize the wall-frame energy
redistribution over the simulated interval.

Figure~\ref{fig:AFB_summary} summarizes the front--behind redistribution of
$\rhodyn$ as an ensemble observable. In the decoupled limit, the dynamical
$s$-field energy is approximately symmetric around the wall, with
$\vev{\AFB}=0.020\pm0.030$. At weak portal coupling, however, the asymmetry
becomes negative:
\begin{equation}
  \vev{\AFB}
  =
  -0.195\pm0.027
  \quad\text{and}\quad
  -0.342\pm0.028
  \label{eq:AFB_weak}
\end{equation}
for $\lambda_{\phi s}=0.25$ and $0.5$, respectively. The excess dynamical
energy is therefore preferentially localized behind the wall in this regime,
while the propagation rate remains close to the decoupled value.

The localization becomes more realization dependent at intermediate coupling.
At $\lambda_{\phi s}=0.75$, the mean asymmetry remains negative,
$\vev{\AFB}=-0.25\pm0.09$, but its uncertainty increases substantially. At
$\lambda_{\phi s}=1$, the result $\vev{\AFB}=0.05\pm0.14$ is consistent with
an approximately symmetric distribution. The large uncertainty indicates that
this coupling lies near a crossover: depending on the stochastic realization,
the excess dynamical energy can be concentrated on either side of the wall.

For stronger coupling, a statistically clearer front-side bias develops. We
obtain
\begin{equation}
  \vev{\AFB}
  \simeq
  0.31,\ 0.47,\ 0.43,\ 0.42
  \label{eq:AFB_strong}
\end{equation}
for $\lambda_{\phi s}=1.25$, $1.5$, $1.75$, and $2$, respectively. Over the
same coupling range, the ensemble-averaged propagation rate decreases from
approximately $0.55$ to $0.42$. Thus, the strong-slowdown regime coincides with
an excess of dynamical $s$-field energy on the false-vacuum side of the wall.

A physical interpretation of this crossover follows from the local curvature
of the potential in the $s$ direction,
\begin{equation}
  m_{s,\mathrm{lin}}^2(\phi,s)
  =
  \frac{\partial^2 V}{\partial s^2}
  =
  m_s^2(T_n)+3\lambda_s s^2
  +\frac{\lambda_{\phi s}}{2}\phi^2.
  \label{eq:s_linearized_mass}
\end{equation}
The portal contribution therefore changes across the wall by
\begin{equation}
  \Delta m_{s,\mathrm{portal}}^2
  =
  \frac{\lambda_{\phi s}}{2}
  \left(\phi_{\rm true}^2-\phi_{\rm false}^2\right)>0.
  \label{eq:s_portal_mass_jump}
\end{equation}
At weak coupling this mass contrast is small. The wall can pass through the
inhomogeneous background with relatively little delay, while the change in the
local $s$-mode frequency excites kinetic and gradient energy that is left in
the wall's wake. This naturally produces the negative asymmetry observed in
the weak-coupling regime.

At stronger coupling, the same inhomogeneous regions with large $s^2$ both
produce a larger portal-induced mass contrast and reduce the local driving
pressure through eq.~\eqref{eq:local_driving_pressure}. The interface then
spends a longer time interacting with those regions. During this delayed
passage, the time-dependent effective mass and the associated spatial
gradients generate and retain dynamical $s$-field energy close to the
false-vacuum side of the wall, providing a natural explanation for the positive
front-side excess. In a mode-scattering description this behavior may also
involve reduced penetration into the higher-mass phase or partial reflection,
but the present observables do not separately determine transmission and
reflection coefficients.

The right panel displays the relation between the two ensemble observables
directly. Across the coupling scan, smaller values of $\vev{\vprop}$ are
generally associated with larger values of $\vev{\AFB}$. As a descriptive
measure of this trend across the nine coupling values, we compute the Pearson
linear correlation coefficient \cite{Pearson:1895},
\begin{equation}
  r_{\rm P}
  =
  \frac{
    \sum_i (A_i-\langle A\rangle)(v_i-\langle v\rangle)
  }{
    \sqrt{
      \sum_i (A_i-\langle A\rangle)^2
      \sum_i (v_i-\langle v\rangle)^2
    }
  },
  \label{eq:pearson_coefficient}
\end{equation}
where $A_i$ and $v_i$ denote the ensemble-averaged front--behind asymmetry and
propagation rate at the $i$th coupling value, respectively. We obtain
$r_{\rm P}\simeq-0.92$. The relation is not strictly monotonic, especially near
the intermediate-coupling crossover where the realization dependence is
largest, but the transition from a behind-the-wall bias at weak coupling to a
front-side bias at strong coupling is clear.

This correlation should not be interpreted as proving that the front-side
excess is the unique cause of the slowdown, because both observables vary with
$\lambda_{\phi s}$. Rather, $\AFB$ provides a complementary wall-frame
diagnostic: the same portal interaction that suppresses the local driving
pressure and reduces the finite-time propagation rate also reorganizes the
dynamical $s$-field energy toward the region into which the wall is advancing.

We emphasize that $\AFB$ is used only as a simple quantitative measure of the front--behind energy redistribution. The physical interpretation does not rely on the precise numerical value of this scalar estimator, but on the corresponding change in the wall-frame distribution of $\rho_s^{\rm dyn}(t,\xi)$.

\begingroup
As a cross-check, we repeated the front--behind analysis including the kinetic
and gradient contributions of both scalar fields,
\begin{equation}
  \rho_{\phi+s}^{\rm dyn}
  =
  \frac{1}{2}\dot{\phi}^{\,2}
  +\frac{1}{2}(\partial_x\phi)^2
  +\frac{1}{2}\dot{s}^{\,2}
  +\frac{1}{2}(\partial_x s)^2 .
  \label{eq:combined_dynamic_energy}
\end{equation}
Although this changes the numerical magnitude of the asymmetry, particularly
at weak coupling, the qualitative transition from a behind-the-wall bias to a
front-side excess with increasing $\lambda_{\phi s}$ is unchanged. The
$\phi$ contribution in Eq.~\eqref{eq:combined_dynamic_energy} contains both
dynamically generated excitations and the kinetic and gradient energy of the
coherent bubble wall. We therefore retain the spectator-sector $\AFB$ as the
primary measure of the additional backreaction channel and use the combined
$(\phi+s)$ result as a consistency check.
\endgroup

\section{Summary and discussion}
\label{sec:summary}

The mechanism studied here differs from conventional plasma friction. No
Boltzmann equation or relativistic fluid is evolved, and no phenomenological
damping term is inserted into the $\phi$ equation. Instead, the coupled scalar
dynamics redistributes energy while conserving the total Hamiltonian energy up
to discretization error. The effective obstruction arises because the
inhomogeneous $s$ background changes the local free-energy difference and can
store kinetic and gradient energy close to the wall.

The present calculation therefore should not be interpreted as a complete
determination of the physical wall velocity in a thermal plasma. In a realistic
model, the scalar-field backreaction considered here can coexist with particle
transport, hydrodynamic heating, and other dissipative effects. The result is
best viewed as an additional contribution to wall backreaction whose importance
depends on the portal coupling and the amplitude and correlation structure of
the classical $s$ background.

The asymmetric initial condition is also deliberately limited in scope. It
isolates the effect of the additional field and can approximate histories in
which $s$ retains much larger classical inhomogeneities than $\phi$. We do not
derive this hierarchy or the chosen finite-cutoff spectrum from a specific
pre-transition model. A future extension could evolve a concrete microscopic
scenario and match its resulting classical spectrum onto the phase-transition
simulation.

Sufficiently large spectator fluctuations may also modify the nucleation
history through their effect on the local effective potential. The present
calculation conditions on an already nucleated bubble and isolates its
subsequent propagation; a complete cosmological treatment would have to
determine the modified nucleation history and wall evolution consistently.

The dimensional split has a similarly specific purpose. The three-dimensional
runs establish the mechanism for spherical geometry, whereas the
one-dimensional runs provide statistics and convergence tests. Quantitative
claims about the coupling dependence should therefore be based on the
one-dimensional ensemble results, with the three-dimensional simulations used
to demonstrate geometric robustness. High-statistics three-dimensional scans
would be valuable but are not required for the proof-of-principle separation of
these roles.
Finally, the wall-frame asymmetry describes scalar-field energy redistribution,
not an exact fluid deflagration or detonation solution. Nevertheless, the sign
of $\AFB$ is physically informative: a negative value records excitations left
in the wall's wake, whereas a persistent positive value identifies excess
energy on the side into which the wall is advancing. This distinction may be
relevant to the partition of transition energy and consequently to predictions
of gravitational waves and baryogenesis that depend on wall propagation.

In summary, we have formulated a post-nucleation initial-value problem for
bubble-wall propagation in a two-scalar-field model. The $\phi$ field forms a
prescribed bubble, and an additional scalar $s$ carries a classical stochastic
background. The asymmetric initialization is used to isolate the $s$-field
backreaction and can approximate cosmological histories satisfying the
hierarchy in eq.~\eqref{eq:fluctuation_hierarchy} at the onset of the
transition.
The portal interaction makes the driving pressure spatially dependent.
Representative three-dimensional simulations show intermittent slowdown in a
spherical geometry. One-dimensional simulations provide the quantitative
framework for measuring the finite-time averaged propagation rate, its portal-coupling dependence, and the statistical variation across stochastic
realizations.
We have also introduced wall-frame observables that determine whether the
kinetic and gradient energy of the additional field is deposited behind the
wall or accumulated on its false-vacuum side. These diagnostics distinguish a
localized two-field backreaction from a uniform damping prescription without
identifying the scalar-energy distribution with an exact hydrodynamic reaction front.
The multi-seed coupling scan, fluctuation-amplitude scan,
energy-conservation test, and fixed-cutoff lattice-refinement study show that
the observed suppression is not attributable to sampling noise or numerical
energy loss. All quoted results are restricted to times before contamination
from periodic images. A continuum extrapolation and a high-statistics
three-dimensional finite-volume study are left for future work. Within these
limitations, the framework provides a direct test of fluctuation-induced scalar
backreaction as an additional ingredient in cosmological bubble-wall dynamics.

\appendix
\section{Lattice implementation details}
\label{app:numerical}

\subsection{Details of the lattice algorithm}
\label{app:algorithm}

Let $q$ denote either scalar field and let $\pi_q=\dot q$ be its canonical
momentum. A staggered leapfrog step may be written schematically as
\begin{align}
\pi_q^{n+1/2}&=\pi_q^{n-1/2}
+\Delta t\left(\nabla^2_{\rm lat}q^n
-\left.\frac{\partial V}{\partial q}\right|_n\right),
\label{eq:leapfrog_momentum}\\
q^{n+1}&=q^n+\Delta t\,\pi_q^{n+1/2}.
\label{eq:leapfrog_field}
\end{align}
The lattice Laplacian is evaluated with centered second-order differences. The
initial half-step momentum is constructed consistently from the specified
$q(0)$ and $\dot q(0)$. When evaluating the energy, the staggered momenta are
interpolated back to the same integer time level as the fields. This avoids
introducing an artificial high-frequency energy oscillation from comparing
quantities defined at different leapfrog time levels.

\section{Numerical validation}
\label{app:numerical_validation}

The slowdown reported in the main text must be distinguished from numerical
energy loss and from residual lattice-spacing dependence.
In addition to energy conservation and the resolution dependence
of the finite-time propagation rate, we also examine timestep sensitivity and
the stability of the late-time propagation behavior in a larger physical
volume.

\paragraph{Energy conservation.}

\begin{figure}[t]
\centering
\safeincludegraphics[width=0.98\textwidth]{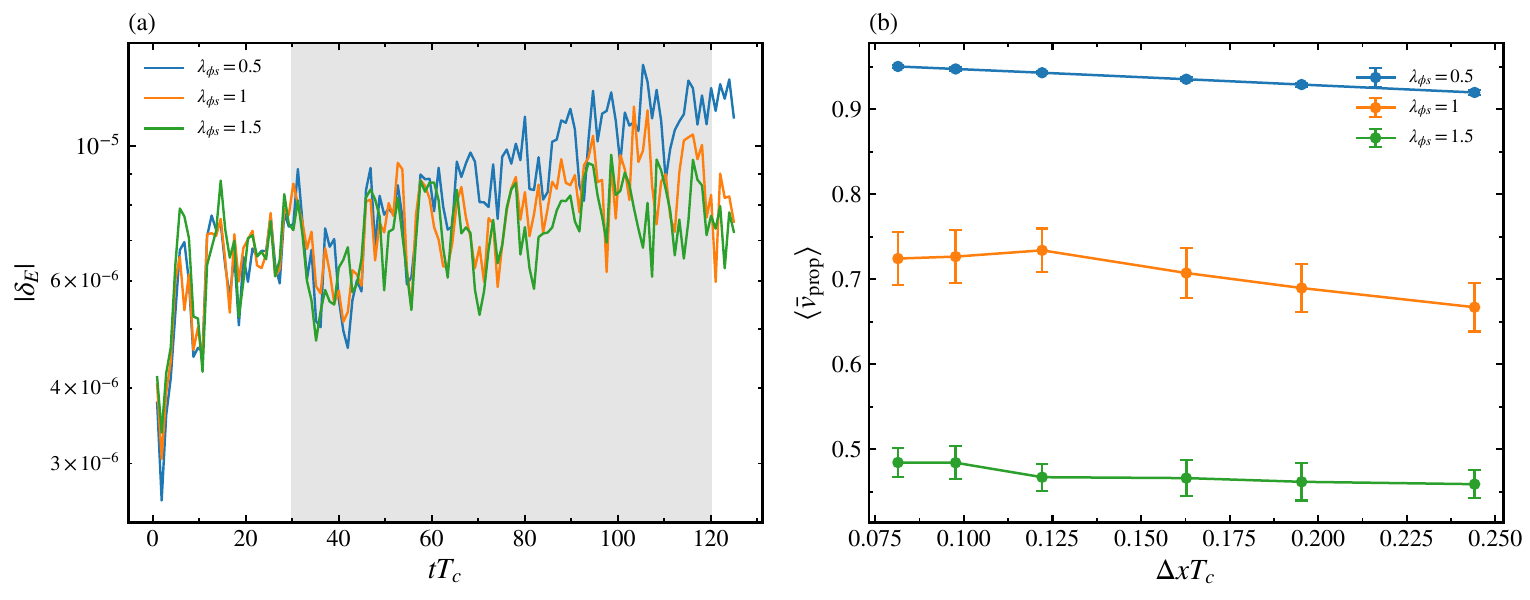}
\caption{Numerical validation of the one-dimensional simulations.
(a) Relative energy drift for representative couplings at $N=2048$; the
shaded region indicates the analysis interval $30\le tT_c\le120$.
(b) Ensemble-averaged finite-time propagation rate versus lattice spacing at
fixed physical $k_{\rm cut}$. Error bars denote the standard error of the mean
over twenty stochastic realizations.}
\label{fig:numerical_validation_appendix}
\end{figure}

We monitor the relative total-energy drift
\begin{equation}
  \delta_E(t)=\frac{E(t)-E(0)}{E(0)},
  \qquad
  E(t)=\int \dd x\,\rho_{\rm tot}(t,x).
  \label{eq:energy_error_appendix}
\end{equation}
Panel~(a) of figure~\ref{fig:numerical_validation_appendix} shows
$|\delta_E|$ for representative couplings at the production resolution
$N=2048$. The point at $t=0$ is omitted, since $\delta_E(0)=0$ by definition
and is not informative on a logarithmic scale. Over the analysis interval
$30\le tT_c\le120$, the maximum drift remains at the level of
$\mathcal{O}(10^{-5})$, with the largest value below approximately
$1.4\times10^{-5}$. The drift is therefore far too small to account for the
observed reduction of the propagation rate.

\paragraph{Resolution dependence.}

To test the sensitivity to the lattice spacing, we compare simulations with
$N=2048$, $2560$, $3072$, $4096$, $5120$, and $6144$ at fixed $LT_c=500$.
The physical stochastic cutoff is held fixed, so that changing $N$ modifies the
discretization without changing the physical ultraviolet content of the initial
ensemble. For each resolution and each portal coupling, we average over twenty
independent stochastic realizations.

Panel~(b) of figure~\ref{fig:numerical_validation_appendix} shows the
ensemble-averaged finite-time propagation rate as a function of
$\Delta xT_c$. The portal-coupling ordering is stable across the full scan,
with
\begin{equation}
  \vev{\vprop}_{\lambda_{\phi s}=0.5}
  >
  \vev{\vprop}_{\lambda_{\phi s}=1}
  >
  \vev{\vprop}_{\lambda_{\phi s}=1.5}
\end{equation}
at every resolution. The two finest lattices already agree closely:
\begin{align}
  \vev{\vprop}(N=5120)
  &= (0.9475\pm0.0020,\ 0.7268\pm0.0308,\ 0.4843\pm0.0194),
  \nonumber\\
  \vev{\vprop}(N=6144)
  &= (0.9504\pm0.0020,\ 0.7245\pm0.0307,\ 0.4846\pm0.0174),
  \label{eq:two_finest_resolutions}
\end{align}
for $\lambda_{\phi s}=(0.5,1,1.5)$, respectively, where the uncertainties are
standard errors over twenty realizations. The corresponding relative shifts
between $N=5120$ and $6144$ are approximately $0.31\%$, $0.32\%$, and
$0.06\%$. Thus, the finite-time propagation rates are stable at the
sub-percent level on the finest lattices. A small residual monotonic trend is
still visible for $\lambda_{\phi s}=0.5$, so we do not attempt a continuum
extrapolation.

\paragraph{Timestep sensitivity.}

{
We have also verified that the extracted propagation rate is insensitive to the
smaller timestep adopted in the $1+1$-dimensional production runs. For
$N=2048$, $LT_c=500$, and $\lambda_{\phi s}=0.8$, increasing the timestep by a
factor of ten from $\Delta t=0.025\,\Delta x$ to
$\Delta t=0.25\,\Delta x$ changes the fitted propagation rate over
$30\leq tT_c\leq120$ from $0.7915$ to $0.8007$, while
$R_{\rm lin}^2\simeq0.997$ in both cases. Repeating this comparison for three
independent stochastic realizations gives relative shifts of approximately
$1.16\%$, $0.38\%$, and $0.12\%$. The larger timestep remains numerically
stable, whereas the smaller timestep reduces the maximum relative energy drift
from the $10^{-3}$ level to the $10^{-5}$ level. We therefore retain
$\Delta t=0.025\,\Delta x$ as a conservative accuracy choice rather than as a
stability requirement.
}

\paragraph{Long-time evolution.}

{
We further tested the duration of the $1+1$-dimensional evolution by increasing
the physical volume to $N=3072$ and $LT_c=750$ while keeping
$\Delta x\,T_c=0.244140625$ fixed, and extending the evolution to $tT_c=300$.
For $\lambda_{\phi s}=0.5$, fits over the consecutive intervals
$30\leq tT_c\leq120$, $120\leq tT_c\leq210$, and
$210\leq tT_c\leq300$ give
\begin{equation}
  \bar v_{\rm prop}=0.9627,\qquad 0.9609,\qquad 0.9643,
\end{equation}
with $R_{\rm lin}^2\simeq1$ in all three intervals. Thus, the approximately
stationary weak-coupling propagation observed in the production window persists
to substantially later times. At stronger coupling, the extended trajectories
continue to exhibit slowing, temporary blocking, and renewed expansion,
indicating that the intermittent behavior is not caused by the finite duration
of the original runs.
}

These tests support the interpretation that the observed slowdown is physical.
The total-energy drift is negligible compared with the size of the effect, and
the coupling dependence of the propagation rate remains stable under lattice
refinement. The strongest quantitative statements in the main text should,
however, be interpreted as statements about the finite-time propagation rate at
high but finite resolution rather than as a strict continuum-limit result.

\acknowledgments

We thank Jing Liu for helpful discussions. This work is supported in part by
the National Natural Science Foundation of China under Grants No.~12235019 and
No.~12475067.

\bibliographystyle{JHEP}
\bibliography{references}

@article{LythWands2002,
  author  = {Lyth, David H. and Wands, David},
  title   = {Generating the curvature perturbation without an inflaton},
  journal = {Phys. Lett. B},
  volume  = {524},
  pages   = {5--14},
  year    = {2002},
  doi     = {10.1016/S0370-2693(01)01366-1},
  eprint  = {hep-ph/0110002},
  archivePrefix = {arXiv},
  primaryClass  = {hep-ph}
}

@article{Greene1997,
  author  = {Greene, Patrick B. and Kofman, Lev and Linde, Andrei and Starobinsky, Alexei A.},
  title   = {Structure of resonance in preheating after inflation},
  journal = {Phys. Rev. D},
  volume  = {56},
  pages   = {6175--6192},
  year    = {1997},
  doi     = {10.1103/PhysRevD.56.6175},
  eprint  = {hep-ph/9705347},
  archivePrefix = {arXiv},
  primaryClass  = {hep-ph}
}

@article{Khlebnikov1998,
  author  = {Khlebnikov, S. and Kofman, L. and Linde, A. and Tkachev, I.},
  title   = {First-order nonthermal phase transition after preheating},
  journal = {Phys. Rev. Lett.},
  volume  = {81},
  pages   = {2012--2015},
  year    = {1998},
  doi     = {10.1103/PhysRevLett.81.2012},
  eprint  = {hep-ph/9804425},
  archivePrefix = {arXiv},
  primaryClass  = {hep-ph}
}

@article{Antusch2023,
  author  = {Antusch, Stefan and Marschall, Kenneth and Torrenti, Francisco},
  title   = {Characterizing the post-inflationary reheating history. Part II: Multiple interacting daughter fields},
  journal = {JCAP},
  volume  = {02},
  pages   = {019},
  year    = {2023},
  doi     = {10.1088/1475-7516/2023/02/019},
  eprint  = {2206.06319},
  archivePrefix = {arXiv},
  primaryClass  = {astro-ph.CO}
}

@article{Aarts:2001yn,
    author        = {Aarts, Gert and Berges, J{\"u}rgen},
    title         = {Classical Aspects of Quantum Fields Far from Equilibrium},
    journal       = {Phys. Rev. Lett.},
    volume        = {88},
    pages         = {041603},
    year          = {2002},
    eprint        = {hep-ph/0107129},
    archivePrefix = {arXiv},
    doi           = {10.1103/PhysRevLett.88.041603}
}

@article{Berges:2004yj,
    author        = {Berges, J{\"u}rgen},
    title         = {Introduction to Nonequilibrium Quantum Field Theory},
    journal       = {AIP Conf. Proc.},
    volume        = {739},
    number        = {1},
    pages         = {3--62},
    year          = {2004},
    eprint        = {hep-ph/0409233},
    archivePrefix = {arXiv},
    primaryClass  = {hep-ph},
    doi           = {10.1063/1.1843591}
}

@article{Berges:2013fga,
    author        = {Berges, J{\"u}rgen and Boguslavski, Kirill
                     and Schlichting, S{\"o}ren and Venugopalan, Raju},
    title         = {Basin of Attraction for Turbulent Thermalization
                     and the Range of Validity of Classical-Statistical Simulations},
    journal       = {JHEP},
    volume        = {05},
    pages         = {054},
    year          = {2014},
    eprint        = {1312.5216},
    archivePrefix = {arXiv},
    primaryClass  = {hep-ph},
    doi           = {10.1007/JHEP05(2014)054}
}

@article{Pearson:1895,
    author  = {Pearson, Karl},
    title   = {Note on Regression and Inheritance in the Case of Two Parents},
    journal = {Proceedings of the Royal Society of London},
    volume  = {58},
    number  = {347--352},
    pages   = {240--242},
    year    = {1895},
    doi     = {10.1098/rspl.1895.0041}
}

@article{LISA:2017pwj,
  author        = {Amaro-Seoane, Pau and others},
  title         = {{Laser Interferometer Space Antenna}},
  eprint        = {1702.00786},
  archivePrefix = {arXiv},
  primaryClass  = {astro-ph.IM},
  year          = {2017}
}

@article{LISA:2024hlh,
  author        = {Colpi, Monica and others},
  title         = {{LISA Definition Study Report}},
  eprint        = {2402.07571},
  archivePrefix = {arXiv},
  primaryClass  = {astro-ph.CO},
  year          = {2024}
}

@article{Hu:2017mde,
  author  = {Hu, Wen-Rui and Wu, Yue-Liang},
  title   = {{The Taiji Program in Space for gravitational wave physics and the nature of gravity}},
  journal = {Natl. Sci. Rev.},
  volume  = {4},
  number  = {5},
  pages   = {685--686},
  year    = {2017},
  doi     = {10.1093/nsr/nwx116}
}

@article{Luo:2021qji,
  author  = {Luo, Ziren and Wang, Yan and Wu, Yueliang and Hu, Wenrui and Jin, Gang},
  title   = {{The Taiji program: A concise overview}},
  journal = {Prog. Theor. Exp. Phys.},
  volume  = {2021},
  number  = {5},
  pages   = {05A108},
  year    = {2021},
  doi     = {10.1093/ptep/ptaa083}
}

@article{TianQin:2020hid,
  author        = {Mei, Jianwei and others},
  title         = {{The TianQin project: current progress on science and technology}},
  journal       = {Prog. Theor. Exp. Phys.},
  volume        = {2021},
  number        = {5},
  pages         = {05A107},
  year          = {2021},
  eprint        = {2008.10332},
  archivePrefix = {arXiv},
  primaryClass  = {gr-qc},
  doi           = {10.1093/ptep/ptaa114}
}

@article{TianQin:2015yph,
  author        = {Luo, Jun and others},
  title         = {{TianQin: a space-borne gravitational wave detector}},
  journal       = {Class. Quant. Grav.},
  volume        = {33},
  number        = {3},
  pages         = {035010},
  year          = {2016},
  eprint        = {1512.02076},
  archivePrefix = {arXiv},
  primaryClass  = {astro-ph.IM},
  doi           = {10.1088/0264-9381/33/3/035010}
}

@article{Hindmarsh:2013xza,
  author        = {Hindmarsh, Mark and Huber, Stephan J. and Rummukainen, Kari and Weir, David J.},
  title         = {{Numerical simulations of acoustically generated gravitational waves at a first order phase transition}},
  journal       = {Phys. Rev. Lett.},
  volume        = {112},
  pages         = {041301},
  year          = {2014},
  eprint        = {1304.2433},
  archivePrefix = {arXiv},
  primaryClass  = {hep-ph},
  doi           = {10.1103/PhysRevLett.112.041301}
}

@article{Espinosa:2010hh,
  author        = {Espinosa, Jose R. and Konstandin, Thomas and No, Jose M. and Servant, Geraldine},
  title         = {{Energy Budget of Cosmological First-order Phase Transitions}},
  journal       = {JCAP},
  volume        = {06},
  pages         = {028},
  year          = {2010},
  eprint        = {1004.4187},
  archivePrefix = {arXiv},
  primaryClass  = {hep-ph},
  doi           = {10.1088/1475-7516/2010/06/028}
}

@article{Caprini:2019egz,
  author        = {Caprini, Chiara and others},
  title         = {{Detecting gravitational waves from cosmological phase transitions with LISA: an update}},
  journal       = {JCAP},
  volume        = {03},
  pages         = {024},
  year          = {2020},
  eprint        = {1910.13125},
  archivePrefix = {arXiv},
  primaryClass  = {astro-ph.CO},
  doi           = {10.1088/1475-7516/2020/03/024}
}

@article{Morrissey:2012db,
  author        = {Morrissey, David E. and Ramsey-Musolf, Michael J.},
  title         = {{Electroweak baryogenesis}},
  journal       = {New J. Phys.},
  volume        = {14},
  pages         = {125003},
  year          = {2012},
  eprint        = {1206.2942},
  archivePrefix = {arXiv},
  primaryClass  = {hep-ph},
  doi           = {10.1088/1367-2630/14/12/125003}
}

@article{Cline:2020jre,
  author        = {Cline, James M. and Kainulainen, Kimmo},
  title         = {{Electroweak baryogenesis at high wall velocities}},
  journal       = {Phys. Rev. D},
  volume        = {101},
  number        = {6},
  pages         = {063525},
  year          = {2020},
  eprint        = {2001.00568},
  archivePrefix = {arXiv},
  primaryClass  = {hep-ph},
  doi           = {10.1103/PhysRevD.101.063525}
}

@article{Wang:2020jrd,
  author        = {Wang, Xiao and Huang, Fa Peng and Zhang, Xinmin},
  title         = {{Phase transition dynamics and gravitational wave spectra of strong first-order phase transition in supercooled universe}},
  journal       = {JCAP},
  volume        = {05},
  pages         = {045},
  year          = {2020},
  eprint        = {2003.08892},
  archivePrefix = {arXiv},
  primaryClass  = {hep-ph},
  doi           = {10.1088/1475-7516/2020/05/045}
}

@article{Bodeker:2009qy,
  author        = {Bodeker, Dietrich and Moore, Guy D.},
  title         = {{Can electroweak bubble walls run away?}},
  journal       = {JCAP},
  volume        = {05},
  pages         = {009},
  year          = {2009},
  eprint        = {0903.4099},
  archivePrefix = {arXiv},
  primaryClass  = {hep-ph},
  doi           = {10.1088/1475-7516/2009/05/009}
}

@article{Leitao:2015fmj,
  author        = {Leitao, Leonardo and Megevand, Ariel},
  title         = {{Gravitational waves from a very strong electroweak phase transition}},
  journal       = {JCAP},
  volume        = {05},
  pages         = {037},
  year          = {2016},
  eprint        = {1512.08962},
  archivePrefix = {arXiv},
  primaryClass  = {astro-ph.CO},
  doi           = {10.1088/1475-7516/2016/05/037}
}

@article{Moore:1995ua,
  author        = {Moore, Guy D. and Prokopec, Tomislav},
  title         = {{Bubble wall velocity in a first order electroweak phase transition}},
  journal       = {Phys. Rev. Lett.},
  volume        = {75},
  pages         = {777--780},
  year          = {1995},
  eprint        = {hep-ph/9503296},
  archivePrefix = {arXiv},
  doi           = {10.1103/PhysRevLett.75.777}
}

@article{Konstandin:2010cd,
  author        = {Konstandin, Thomas and No, Jose M.},
  title         = {{Hydrodynamic obstruction to bubble expansion}},
  journal       = {JCAP},
  volume        = {02},
  pages         = {008},
  year          = {2011},
  eprint        = {1011.3735},
  archivePrefix = {arXiv},
  primaryClass  = {hep-ph},
  doi           = {10.1088/1475-7516/2011/02/008}
}

@article{Krajewski:2024gma,
  author        = {Krajewski, Tomasz and Lewicki, Marek and Zych, Mateusz},
  title         = {{Bubble-wall velocity in local thermal equilibrium: hydrodynamical simulations vs analytical treatment}},
  journal       = {JHEP},
  volume        = {05},
  pages         = {011},
  year          = {2024},
  eprint        = {2402.15408},
  archivePrefix = {arXiv},
  primaryClass  = {astro-ph.CO},
  doi           = {10.1007/JHEP05(2024)011}
}

@article{Kurki-Suonio:1995yaf,
  author        = {Kurki-Suonio, Hannu and Laine, Mikko},
  title         = {{On bubble growth and droplet decay in cosmological phase transitions}},
  journal       = {Phys. Rev. D},
  volume        = {54},
  pages         = {7163--7171},
  year          = {1996},
  eprint        = {hep-ph/9512202},
  archivePrefix = {arXiv},
  doi           = {10.1103/PhysRevD.54.7163}
}

@article{Ignatius:1993qn,
  author        = {Ignatius, Jan and Kajantie, Keijo and Kurki-Suonio, Hannu and Laine, Mikko},
  title         = {{The growth of bubbles in cosmological phase transitions}},
  journal       = {Phys. Rev. D},
  volume        = {49},
  pages         = {3854--3868},
  year          = {1994},
  eprint        = {astro-ph/9309059},
  archivePrefix = {arXiv},
  doi           = {10.1103/PhysRevD.49.3854}
}

@inproceedings{Quiros:1999jp,
  author        = {Quiros, Mariano},
  title         = {{Finite temperature field theory and phase transitions}},
  booktitle     = {{High Energy Physics and Cosmology: Proceedings of the Summer School, Trieste, Italy, 29 June--17 July 1998}},
  pages         = {187--259},
  year          = {1999},
  eprint        = {hep-ph/9901312},
  archivePrefix = {arXiv},
  reportNumber  = {IEM-FT-187-99}
}

@article{Profumo:2014opa,
  author        = {Profumo, Stefano and Ramsey-Musolf, Michael J. and Wainwright, Carroll L. and Winslow, Peter},
  title         = {{Singlet-catalyzed electroweak phase transitions and precision Higgs boson studies}},
  journal       = {Phys. Rev. D},
  volume        = {91},
  number        = {3},
  pages         = {035018},
  year          = {2015},
  eprint        = {1407.5342},
  archivePrefix = {arXiv},
  primaryClass  = {hep-ph},
  doi           = {10.1103/PhysRevD.91.035018}
}

@article{Tofighi:2015fia,
  author        = {Tofighi, Ali and Ghodsi, Omid N. and Saeedhoseini, Mehrdad},
  title         = {{Phase transition in the singlet scalar dark matter model}},
  journal       = {Phys. Lett. B},
  volume        = {748},
  pages         = {208--211},
  year          = {2015},
  eprint        = {1510.00791},
  archivePrefix = {arXiv},
  primaryClass  = {hep-ph},
  doi           = {10.1016/j.physletb.2015.07.009}
}

@article{Zhou:2020ojf,
  author        = {Zhou, Ruiyu and Yang, Jing and Bian, Ligong},
  title         = {{Gravitational waves from first-order phase transition and domain wall}},
  journal       = {JHEP},
  volume        = {04},
  pages         = {071},
  year          = {2020},
  eprint        = {2001.04741},
  archivePrefix = {arXiv},
  primaryClass  = {hep-ph},
  doi           = {10.1007/JHEP04(2020)071}
}

@article{Li:2025zxa,
  author        = {Li, Yuan-Jie and Liu, Jing and Guo, Zong-Kuan},
  title         = {{Enhancement of first-order phase transitions through a mass-acquiring scalar field}},
  eprint        = {2508.14665},
  archivePrefix = {arXiv},
  primaryClass  = {hep-ph},
  year          = {2025}
}

\end{document}